\documentclass[%
 reprint,nofootinbib,
 amsmath,amssymb,
 aps,
]{revtex4-1}

\usepackage[left]{lineno}

\usepackage{graphicx}
\usepackage{dcolumn}
\usepackage{bm}
\usepackage{hyperref}
\hypersetup{
    colorlinks=true,
    linkcolor=blue,
    filecolor=blue,      
    urlcolor=blue,
    citecolor=blue
    }
\usepackage{amsmath,amssymb,amsfonts,bm}
\usepackage{ulem}
\usepackage[dvipsnames]{xcolor}

\begin{document}


\title{Probing the sensitivity of anisotropic flow coefficients to the initial nuclear structure in pO and OO collisions at the LHC}



\author{Aswathy Menon Kavumpadikkal Radhakrishnan$^{1,3}$}\email[]{Aswathy.Menon@cern.ch}
\author{Suraj Prasad$^{1,3}$}\email[]{Suraj.Prasad@cern.ch}
\author{Neelkamal Mallick$^2$}\email[]{Neelkamal.Mallick@cern.ch}
\author{Raghunath Sahoo$^1$}\email[Corresponding author:]{Raghunath.Sahoo@cern.ch}
\author{Gergely Gábor Barnaföldi$^3$}\email[]{barnafoldi.gergely@wigner.hun-ren.hu}
\affiliation{$^1$Department of Physics, Indian Institute of Technology Indore, Simrol, Indore 453552, India}
\affiliation{$^2$University of Jyv\"askyl\"a, Department of Physics, P.O. Box 35, FI-40014, Jyv\"askyl\"a, Finland}
\affiliation{$^3$HUN-REN Wigner Research Centre for Physics, 29-33 Konkoly-Thege Miklós Str., H-1121 Budapest, Hungary}

\date{\today} 

\begin{abstract}
RHIC and LHC have injected $^{16}\rm O$ nuclei in their accelerator complexes with a focus on investigating collectivity and the origin of quark-gluon plasma signatures in small collision systems. The $^{16}\rm O$ nuclei are known to possess clusters of $\alpha$-particles ($^{4}\rm He$) inside the nucleus. This paper attempts to study the clustered-nuclear-geometry dependence of anisotropic flow coefficients such as elliptic flow ($v_2$) and triangular flow ($v_3$), which are sensitive to the nuclear geometry of colliding nuclei. The study is performed in pO and OO collisions at $\sqrt{s_{\rm NN}}=9.61$~TeV and 7~TeV respectively, employing a hybrid model encompassing IP-Glasma + MUSIC + iSS + UrQMD. The results of the clustered nuclear geometry are compared with those of the Woods\,--\,Saxon nuclear profile. Both initial and final state anisotropies are estimated. This study is thus one of its first kind, where the study of anisotropic flow coefficients for pO and OO collisions is presented using a hybrid hydrodynamics model. While the effect of $\alpha$-clustering in pO is found to be small, it is significant for each observable studied in OO collisions. It is also observed that the magnitude of this effect has a noteworthy dependence on the size of the \textsuperscript{4}He.

\end{abstract}

\keywords{pO collisions, azimuthal correlations}

\maketitle

\section{INTRODUCTION}
\label{sec:intro}
The heavy-ion collisions at the Large Hadron Collider (LHC), CERN, and the Relativistic Heavy-Ion Collider (RHIC), BNL, aim to recreate a deconfined matter of partons, known to have existed shortly after the Big Bang. This deconfined and thermalised medium of partons, known as Quark--Gluon Plasma (QGP), is transient in nature and is shown to possess collective phenomena similar to fluids~\cite{ALICE:2022wpn, Heinz:2013th}. Due to the short lifetime of QGP, its existence in heavy-ion collisions is often inferred by studying indirect signatures, which include strangeness enhancement~\cite{Rafelski:1982pu}, collectivity~\cite{Voloshin:1994mz}, quarkonia suppression~\cite{Matsui:1986dk}, jet-quenching~\cite{Gyulassy:2000fs, Gyulassy:2000er, Levai:2001dc}, to name a few. Small collision systems like proton-proton (pp) and proton-lead (p--Pb) collisions are often considered as baselines to study QGP and cold nuclear matter effects in heavy-ion collisions. Thanks to the observation of these collective and other heavy-ion-like phenomena in high-multiplicity pp and p--Pb collisions at the LHC energies~\cite{ALICE:2024vzv, ALICE:2016fzo}, the small collision systems are of great interest, where the existence of a QGP medium is hinted at. Thus, both RHIC and LHC prepare to inject small nuclei, like e.g., oxygen ($^{16}\rm O$) ions into the beam pipe to perform oxygen-oxygen (OO) and proton-oxygen (pO) collisions~\cite{Brewer:2021kiv, Katz:2019qwv}. These collision systems are important as they bridge the multiplicity gap between high multiplicity pp, p--Pb, and Pb--Pb collisions. Studying them ought to provide insights into the nature of the QCD medium in small systems where some of the signatures---like jet quenching and quarkonia suppression---are yet to be observed. Further, pO collisions are specifically important as their measurements at the LHC would be useful to tune the cosmic ray air shower models, to solve the existing puzzles in muon production~\cite{Scaria:2023coa}.

One of the key aspects to investigate in pO and OO collisions is rooted in low-energy nuclear physics, which suggests the presence of $\alpha$-clustered nuclear structure in light nuclei having $4n$ number of nucleons. Oxygen ($^{16}\rm O$) is one such nuclei where the $\alpha$-particles ($^{4}\rm He$ nuclei) arrange themselves at the corners of a regular tetrahedron~\cite{gamow, Wheeler:1937zza, Bijker:2014tka, Wang:2019dpl, He:2014iqa, He:2021uko, Otsuka:2022bcf}. In heavy-ion collisions, several studies argue for a modification in the final state particle production due to a variation in the initial state nuclear structure of the colliding nuclei~\cite{Behera:2023nwj, Giacalone:2021udy, Haque:2019vgi, Behera:2023oxe, ALICE:2021ibz, ALICE:2018lao, ATLAS:2019dct, CMS:2019cyz, STAR:2015mki, Giacalone:2021udy, Haque:2019vgi, PHENIX:2018lia}. The geometry scan program at RHIC is dedicated to understanding the particle production due to the specific geometry of the colliding nuclei~\cite{PHENIX:2018lia, PHENIX:2021ubk, STAR:2022pfn, STAR:2023wmd}. A number of studies at the RHIC and LHC energies have been performed to study the impact of clustered structure in the final state particle production and onset of hydrodynamic flow in OO collisions~\cite{Li:2020vrg, Rybczynski:2019adt, Sievert:2019zjr, Huang:2019tgz, Behera:2023nwj, Behera:2021zhi, Lim:2018huo, Summerfield:2021oex, Schenke:2020mbo, Rybczynski:2019adt, Sievert:2019zjr, Huang:2019tgz, Huss:2020whe, Zakharov:2021uza, Giacalone:2024ixe, Zhang:2024vkh, R:2024eni, Prasad:2024ahm, Ding:2023ibq, Wang:2021ghq, Rybczynski:2017nrx, Svetlichnyi:2023nim}. It is observed that due to a compact nuclear geometry, OO collisions with clustered nuclear geometry yield a higher particle multiplicity and energy density in the central collisions as compared to a traditional Woods--Saxon nuclear profile~\cite{Behera:2021zhi}. Additionally, the clustered OO collisions lead to significantly large values of triangular flow~\cite{Behera:2023nwj, Prasad:2024ahm}. One of our earlier studies concluded that the anisotropic flow fluctuations are sensitive to the clustered nuclear geometry of the colliding nuclei~\cite{Prasad:2024ahm}. In several studies, the authors have also performed O--Au, C--Au, O--Pb, and Ne--Pb collisions at the RHIC and LHC energies to investigate the clustered structure of light nuclei~\cite{Bozek:2014cva, Broniowski:2013dia, Giacalone:2024luz, Lim:2018huo}. However, similar studies are underperformed in pO collisions, which are interesting and relevant to the understanding of the effect of the clustered structure of $^{16}\rm O$ in asymmetric small collisions, aiding in identifying the effects arising from the initial nuclear profile and small system dynamics. Additionally, anisotropic flow measurements in pO and OO collisions are capable of making remarkable additions to the present understanding of partonic collectivity recently observed in pp and p--Pb collisions by the ALICE collaboration~\cite{ALICE:2024vzv}. Interestingly, one recent study based on a multi-phase transport (AMPT) model investigates the effect of clustered nuclear geometry in pO collisions~\cite{R:2024eni}, which shows that the effects of $\alpha$-clustering are significantly different from those of the unclustered nuclear profile, as far as the flow-observables under study are concerned. The uniqueness of $\alpha$-cluster results is found to be consistent with that of the AMPT results on OO collisions at $\sqrt{s_{\rm NN}}= 7$~TeV as well~\cite{R:2024eni, Behera:2021zhi}.

In this work, we simulate pO and OO collisions at $\sqrt{s_{\rm NN}}=9.61$ TeV and 7 TeV, respectively, using a hybrid model referred to as IP-Glasma + MUSIC + iSS + UrQMD. Every observable of interest is studied for two nuclear density profiles of $^{16}\rm O$ nuclei---$\alpha$-clustered and Woods--Saxon nuclear profiles. The initial state eccentricity ($\epsilon_2$) and triangularity ($\epsilon_3$) are calculated, and are compared to the final state anisotropic flow coefficients such as elliptic flow ($v_2$) and triangular flow ($v_3$) for both pO and OO collisions. The anisotropic flow coefficients are estimated using the two-particle Q-cumulant method.

The paper is organized as follows. After the brief introduction in Section~\ref{sec:intro}, Section~\ref{sec:EGM} discusses the event generation using the hybrid model and the two-particle Q-cumulant method. In Section~\ref{sec:R&D}, our results are presented with necessary discussions. The study is summarized in Section~\ref{sec:summary} with a brief outlook. Some of the additional findings that validate our results are provided in the Appendix subsections~\ref{AppendixA} and~\ref{AppendixB}.

\section{Event Generation and Methodology}
\label{sec:EGM}

\subsection{IP-Glasma + MUSIC + iSS + UrQMD}

This work employs a hybrid framework---IP-Glasma+MUSIC+iSS+UrQMD model--- to simulate the evolution of the ultra-relativistic pO and OO collisions, as it gives a good description of particle production and flow across small to large collision systems~\cite{McDonald:2016vlt}. Here, the initial conditions of the collisions are described by the impact-parameter-dependent Glasma (IP-Glasma) model, MUSIC handles the hydrodynamic evolution of the fireball, the \texttt{iSS} package does the particlization from the MUSIC hypersurface, and the UrQMD transport model carries out the final-state hadronic interactions. These four stages are briefly discussed below, along with the details of the parameters/settings used in our study.

\subsubsection{IP-Glasma: Pre-equilibrium}
IP-Glasma is a theoretical framework based on Color Glass Condensate (CGC) effective field theory, which simulates the early stage dynamics of gluon fields during relativistic collisions~\cite{Schenke:2012wb, Schenke:2012hg, McLerran:1993ni, McLerran:1993ka}. It is an integration of two frameworks~\cite{McDonald:2016vlt}: the impact-parameter-dependent dipole saturation (IP-Sat) model~\cite{Bartels:2002cj, Kowalski:2003hm}, which handles the initial nuclear configurations, and the Glasma framework~\cite{Krasnitz:1998ns, Krasnitz:1999wc, Krasnitz:2001qu, Krasnitz:2000gz}, which evolves these configurations post-collision till thermalization. The two nuclei, described as two sheets of CGC fields, are boosted and made to collide with each other at $\tau=0$. Further, the classical Yang--Mills equations, which model the glasma evolution, are solved numerically using a lattice approach (lattice size, $L$ = 14, lattice spacing, $a$ = 0.02 fm)~\cite{Schenke:2020mbo}, which generates the initial energy--momentum tensor ($T_{\rm YM}^{\mu\nu}$) of the system. The model accounts for the two essential sources of event-by-event fluctuations: 1) the random spatial distribution of nucleons inside the colliding nuclei, and 2) the fluctuating color charge densities within each nucleon. The color charge fluctuations are described by the IP-Sat model, where the sub-nucleonic fluctuations are modelled as three Gaussian hotspots per nucleon. 

As far as the setup of nuclear density profiles is concerned, the nucleonic positions inside the nucleus are randomly sampled for each event, according to the Woods--Saxon distribution with the parameter settings being the same as those in Ref.~\cite{Prasad:2024ahm}. In addition, since our aim is to understand the effect of $\alpha$-clustering on the flow coefficients, we also sample nucleons into an $\alpha$-clustered geometry such that the four $\alpha$-clusters of the $^{16} \rm O$ nucleus are at the corners of a regular tetrahedron, while the nucleons inside each cluster are distributed according to Woods--Saxon profile. A detailed description of $\alpha$-clustering implementation in oxygen nuclei is provided towards the end of Section ~\ref{sec:ACimplementation}. Finally, the output of IP-Glasma is a fluctuating energy--momentum tensor computed at $\tau_{\rm switch}=0.4~\rm fm$, which is fed to MUSIC simulations for hydrodynamic evolution.

\subsubsection{MUSIC: Hydrodynamics}
Using viscous relativistic hydrodynamics, MUSIC evolves the energy--momentum tensor obtained from IP-Glasma at the thermalization time $\tau_{0} = 0.4$~fm, under the assumption of local thermal equilibrium. The evolution incorporates shear and bulk viscous effects, solving the conservation law $\partial_\mu T^{\mu\nu}$~\cite{Schenke:2010nt, Schenke:2010rr, Paquet:2015lta}. The simulation is performed in a (2+1)D boost-invariant set-up, using an equation-of-state parametrization \texttt{"s95p-v1.2"} that is obtained from the interplay between lattice-QCD and hadron resonance gas~\cite{Huovinen:2009yb}.  A constant shear viscosity to entropy ratio of $\eta/s = 0.12$ is used along with a temperature-dependent bulk viscosity $\zeta/s(T)$ as described in Ref.~\cite{Schenke:2020mbo}. Further, the Kurganov--Tadmor numerical algorithm solves the hydrodynamic equations~\cite{Kurganov:2000ovy, Jeon:2015dfa}. The evolution continues until the local energy density falls to a switching energy density, $\varepsilon_{\rm switch}= 0.18~\rm GeV/fm^{3}$~\cite{Schenke:2020mbo}, at which a freeze-out hypersurface is constructed, marking the end of the hydrodynamic evolution of the fluid.
 
\subsubsection{iSS: Particlization}
The hydrodynamic freeze-out hypersurface data produced by MUSIC acts as the input for the \texttt{iSS} (\texttt{iSpectraSampler})~\cite{Shen:2014vra, Denicol:2018wdp}, which converts the fluid-like medium into hadrons, using the Cooper--Frye formula~\cite{Cooper:1974mv, Dusling:2009df, Schenke:2020mbo}. Particle sampling is done based on the local flow velocity, temperature, and viscous corrections, reflecting the momentum and spatial distributions of particles emitted at freeze-out. \texttt{iSS} allows oversampling; hence, to increase statistical precision without re-running hydrodynamics, a finite number of events can be sampled from each MUSIC hypersurface for pO(OO) collisions at $\sqrt{s_{\rm NN}}=9.61 (7)$~TeV. The total number of hadronic freezeout events performed per IP-Glasma + MUSIC event, $N_{\rm sample}=$~200 in our study.

\subsubsection{UrQMD: Hadronic cascade}
Particles sampled from \texttt{iSS} are propagated using Ultra-relativistic Quantum Molecular Dynamics (UrQMD, version 3.4) microscopic transport model, with its default settings, which performs the final-stage hadronic interactions such as elastic and inelastic hadronic scatterings, resonance decays, strong decays, and baryon-antibaryon annihilations~\cite{Bass:1998ca, Bleicher:1999xi}. This model, which solves the Boltzmann transport equation using Monte Carlo techniques, handles hadrons up to 2.25 GeV in mass and performs a dynamical freeze-out for different particle species~\cite{Schenke:2020mbo}. The model outputs the final-state four momenta and PIDs, which can be stored for subsequent examination and analysis.

\subsection{Two-particle Q-Cumulant method}
\label{cumu}
One of the important signatures of the hydrodynamic behavior of the system formed in relativistic nuclear collisions is given by collectivity. The studies of anisotropic flow coefficients are thus crucial to understand the collective behaviour of the system formed in nuclear collisions. Anisotropic flow is quantified using the coefficients of the Fourier expansion of the azimuthal distribution of the particles in the final state, given as follows~\cite{Voloshin:1994mz},
\begin{equation}
\frac{dN}{d\phi}\propto 1+\sum_{n=1}^{\infty}2v_{n}\cos[n(\phi-\psi_{n})] \ .
\label{eq:fourierexpansion}
\end{equation}
Here, $\phi$ is the azimuthal angle, $\psi_n$ is the $n$\textsuperscript{th} harmonic symmetry plane angle, and $v_n$ are the coefficients of the Fourier expansion, also known as the anisotropic flow coefficients. $v_2$, $v_3$, etc., are respectively called as elliptic flow, triangular flow, etc, and can be calculated as, $v_{n}=\langle\cos[n(\phi-\psi_{n})]\rangle$. However, the estimation of $\psi_n$ is not trivial in experiments; thus, the two-particle Q-cumulant method is used to estimate the flow coefficients~\cite{Bilandzic:2010jr, Zhou:2015iba}. The estimation of $n$\textsuperscript{th} order flow coefficient with Q-cumulant method requires the $n$\textsuperscript{th} order flow vectors ($Q_n$), defined as follows,
\begin{equation}
    Q_{n}=\sum_{j=1}^{M}e^{in\phi_{j}} \ ,
\end{equation}
where $M$ is the number of charged particles in an event.

As mentioned in Section~\ref{sec:EGM}, in this study, a hybrid of IP-Glasma + MUSIC + iSS + UrQMD models is used, where each IP-Glasma + MUSIC event is passed 200 times through iSS + UrQMD. To reduce the random fluctuations arising due to sampling of a finite number of particles, we define the flow vector of a super event ($Q^{\rm SE}_{n}$) as follows~\cite{McDonald:2016vlt},
\begin{equation}
    Q^{\rm SE}_{n}=\sum_{j=1}^{N_{\rm sample}}\sum_{k=1}^{M_{j}}e^{in\phi_{jk}}
    \label{eq:QnSE} \ .
\end{equation}
Here, $M_{j}$ is the multiplicity of the $j$\textsuperscript{th} hadronic sample, and $\phi_{jk}$ is the azimuthal angle of the $k$\textsuperscript{th} particle in this sample. 
To reduce the contribution of nonflow, one can make two super-sub events, $A$ and $B$ respectively, separated by a pseudorapidity gap, $|\Delta\eta|>1$. Corresponding flow vector can be denoted as, $Q^{A}_{n}$ and $Q^{B}_{n}$ with multiplicities $M_A$ and $M_B$ respectively. Consequently, the two-particle correlation is given by,
\begin{equation}
    \langle 2 \rangle _{\Delta \eta} = \frac{Q_{n}^{A} \cdot Q_{n}^{B *}} {M_{A} \cdot M_{B}} \ .
\end{equation}
Here, the superscript `*' represents the complex conjugate. The corresponding two-particle Q-cumulant can be calculated using the following expression.
\begin{equation}
c_{n}\{2, |\Delta\eta|\} = \langle \langle 2 \rangle \rangle _{\Delta\eta} \ ,
\end{equation}
where, $\langle\langle\dots\rangle\rangle$ denotes the average over the super events. Finally, the anisotropic flow coefficients can be calculated using the following expression.
\begin{equation}
v_{n}\{2, |\Delta\eta|\} =\sqrt{c_{n}\{2, |\Delta\eta|\}}  \ .
\end{equation}

\subsection{Estimation of Spatial Geometry from IP-Glasma}
One can quantify the initial state spatial anisotropy of the collision overlap region using eccentricity ($\epsilon_2$), triangularity ($\epsilon_3$), etc., as follows~\cite{Petersen:2010cw, Prasad:2022zbr}, 
\begin{equation}
\epsilon_{n}=\frac{\sqrt{\langle{r^{n}\cos(n\phi_{\text{part}})}\rangle^{2}+\langle{r^{n}\sin(n\phi_{\text{part}})}\rangle^{2}}}{\langle{r^{n}}\rangle}
\label{eq:eccentricity} \ ,
\end{equation}
where $r$ and $\phi_{\rm part}$ denote the radial distance and azimuthal angle of the participant nucleons from the center in polar coordinates. However, with IP-Glasma, it is possible to estimate the spatial anisotropy of the fluid just before the start of the MUSIC hydrodynamics. Hence, by using energy density, $\varepsilon(x,y)$, in each fluid cell of dimension (${\textrm{d}} x, {\textrm{d}}y$) at ($x,y$) from the center ($0,0$), Eq.~\eqref{eq:eccentricity} can be modified as follows~\cite{Schenke:2013aza}.

\begin{widetext}
\begin{equation}
  \epsilon_{n}=\frac{\sqrt{\big(\iint_{A}{r^{n}\varepsilon(x,y)\cos(n\phi_{\rm part})}{\textrm{d}}x\;{\textrm{d}}y\big)^{2}+\big(\iint_{A}{r^{n}\varepsilon(x,y)\sin(n\phi_{\rm part})}{\textrm{d}}x\;{\textrm{d}}y\big)^{2}}}{\iint_A{r^{n}\varepsilon(x,y){\textrm{d}}x\;{\textrm{d}}y}}  
  \label{eq:eccIPG}
\end{equation}
\end{widetext}
Here, the integral runs over all the fluid cell elements in the transverse area $A$. $r=\sqrt{x^2+y^2}$ is the radial distance of the fluid cell and $\phi_{\rm part}=\tan^{-1}\big(\frac{y}{x}\big)$ is the corresponding azimuthal angle. From here onwards, the IP-Glasma results of eccentricity and triangularity presented are those evaluated using the energy-density-weighted definition as in Eq.~\eqref{eq:eccIPG}.

\subsection{$\alpha$-clustering in $^{16} \rm O$ nucleus}

\begin{figure}[ht!]
\centering
\includegraphics[scale=0.44]{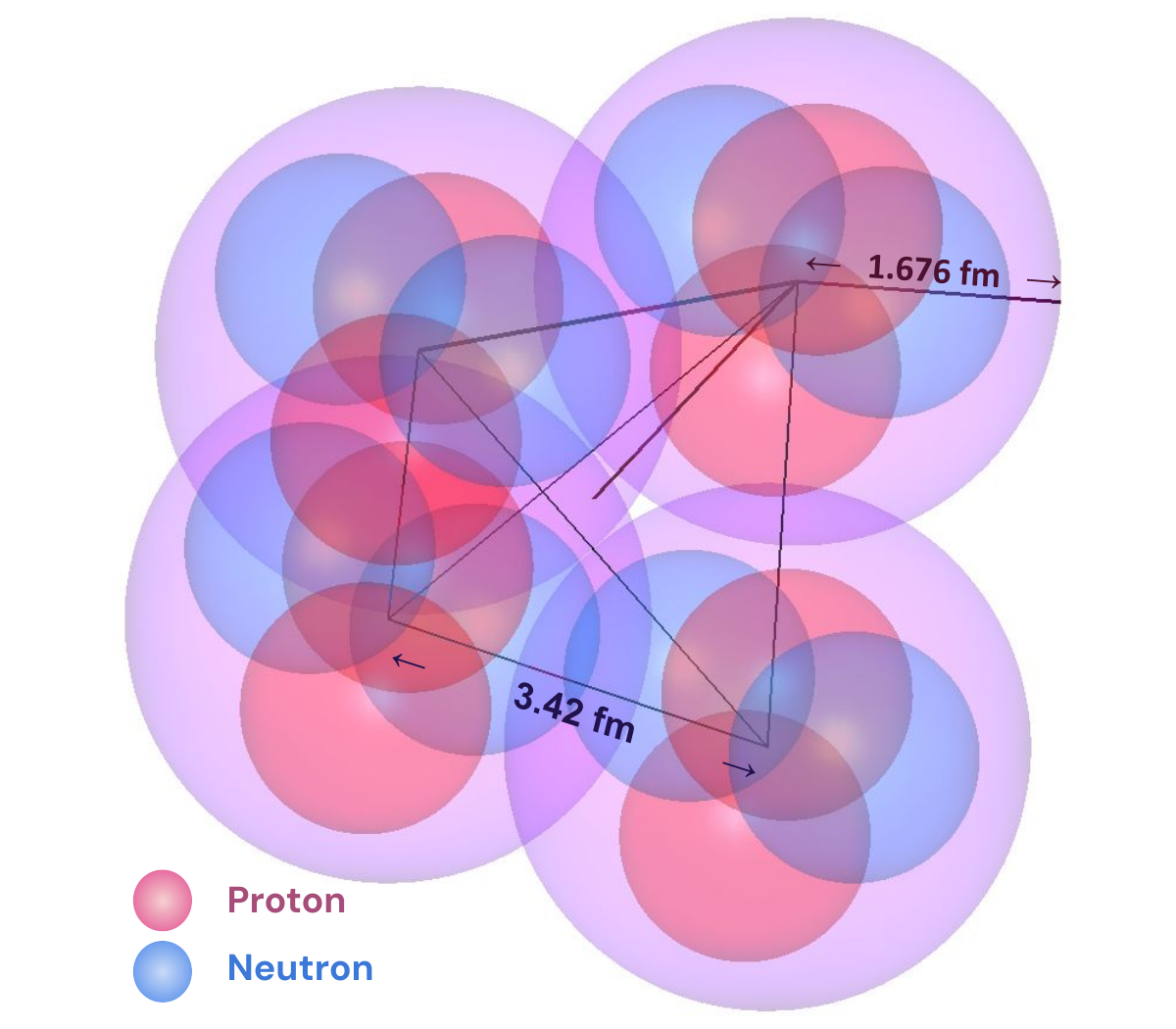}
\caption{Pictorial representation of $\alpha$-clustered $^{16} \rm O$ nucleus (not to scale).}
\label{fig:O}
\end{figure}

\label{sec:ACimplementation}
In self-conjugate light nuclei having $4n$ number of nucleons, such as $^{12}\rm C$, $^{16}\rm O$, $^{20}\rm Ne$, etc., the nucleons are theorized to exist in the form of groups of $\alpha$ particles ($^{4}\rm He$ nuclei). This phenomenon is known as $\alpha$-clustering. Due to the intrinsic stability of $\alpha$-clusters and the fashion in which they arrange themselves inside these nuclei, $\alpha$-clustering contributes to the tight-binding and exceptional stability of these nuclei in comparison to their neighboring nuclei (as evident from the peaks in the binding energy per nucleon curve).

However, observing $\alpha$-clusters is experimentally challenging as the nucleus is never in a static configuration, due to the quantum fluctuations and subnucleonic motions. Thus, what is required are the ``snapshots" of the nuclei for investigating the effects of clustering~\cite{Otsuka:2022bcf, Li:2020vrg}. During relativistic nuclear collisions, due to the very short interaction time compared to the timescales of nuclear dynamics, experimentalists get an opportunity to deal with the frozen-ground-state nuclear configurations. Thus, the high-energy relativistic nuclear collisions at LHC are potential tools to probe the ground-state nuclear structures, especially $\alpha$-clustering as predicted by the low-energy nuclear physics.


The $^{16}\rm O$ is a doubly magic nucleus that consists of 8 protons and 8 neutrons, which can be grouped into four $\alpha$-clusters, each containing two protons and two neutrons as depicted in Fig.~\ref{fig:O}. These four $\alpha$-clusters are arranged at the vertices of a regular tetrahedron. To reproduce the experimental root-mean-squared (RMS) radius of the oxygen nuclei, the side-length of the tetrahedron is set to be 3.42 fm, which yields an RMS radius of 2.699 fm for the $\alpha$-clustered $^{16} \rm O$ nucleus~\cite{Li:2020vrg, Behera:2023nwj, Behera:2021zhi, Prasad:2024ahm, R:2024eni}. The nucleons in each $\alpha$-cluster are sampled according to the Woods--Saxon distribution as :
\begin{equation}
\rho(r) = \frac{\rho_{0} \Big(1+ w \big(\frac{r}{r_{0}}\big)^{2}\Big)}{1 + \exp\big(\frac{r - r_{0}}{a}\big)}, 
\label{eq:WS}
\end{equation}
where $\rho(\rm r)$ and $\rho_{0}$ are the nuclear charge densities at the radial distance $r$ and at $r=0$ respectively, $w=0.517$, is the deformation parameter, $a = 0.322$ fm is the skin-depth and $r_{0}$ = 0.964 fm is the mean radius corresponding to $^{4}\rm He$ nucleus with RMS radius equaling 1.676 fm. To incorporate the short-distance nucleon-nucleon repulsion, the minimum distance of separation between any pair of nucleon centers is fixed to be 0.4 fm~\cite{Loizides:2017ack}. The nucleonic configurations in both projectile and target ($^{16} \rm O$) nuclei are randomized before each collision event.

\section{Results and Discussions}
\label{sec:R&D}

In this section, the generated initial eccentricities via $\epsilon_n$, and measured final state azimuthal anisotropies via $v_n$ ($n=2,~3$) for pO collisions at $\sqrt{s_{\rm NN}}=9.61$ TeV and OO collisions at $\sqrt{s_{\rm NN}}=7$ TeV simulated using the IP-Glasma+MUSIC+iSS+UrQMD model, are presented. The centrality selection is performed via geometrical slicing using the impact-parameter distribution obtained from IP-Glasma. Using the conventional definition of centrality, the most central collisions (\textit{e.g.} 0--10\%) are the almost head-on collisions for which the impact parameter values are close to zero, and the number of participant nucleons, $N_{\rm part}$, is close to the total number of nucleons in both the colliding nuclei.

\begin{figure}[ht!]
\centering
\includegraphics[scale=0.44]{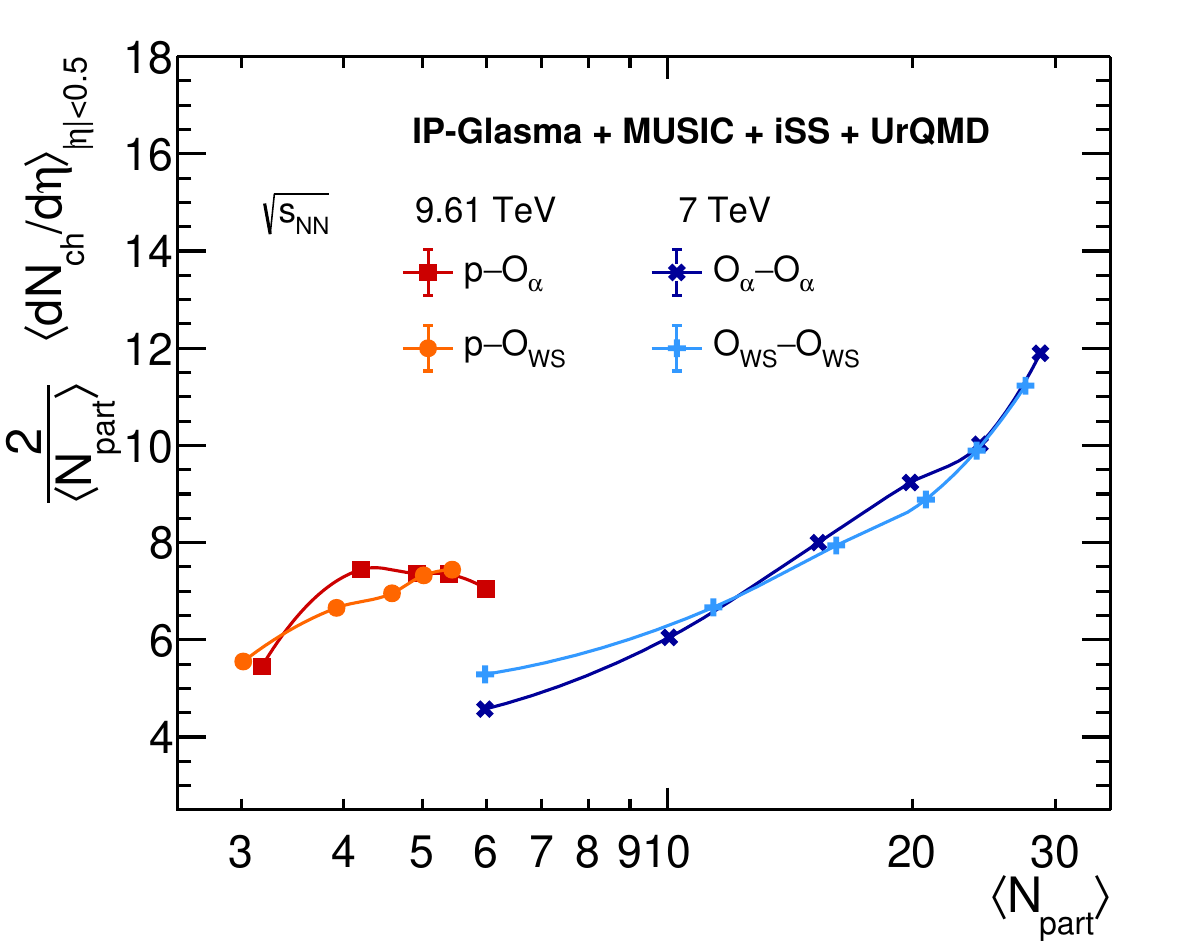}
\caption{Charged particle multiplicity density, $\langle dN_{\rm ch}/d\eta \rangle_{|\eta|<0.5}$, normalized to the average number of participating nucleon pairs, $\langle N_{\rm part} \rangle/2$, plotted as a function of $\langle N_{\rm part} \rangle$ in pO, and OO collisions at $\sqrt{s_{\rm NN}}=9.61$~TeV and $7$~TeV, respectively. Estimated statistical uncertainties are well within the marker sizes.}
\label{fig:dNchdeta}
\end{figure}

Figure~\ref{fig:dNchdeta} shows the charged particle multiplicity density, $\langle dN_{\rm ch}/d\eta \rangle$, at midrapidity, $|\eta|<0.5$, normalized to the average number of participating nucleon pairs, $\langle N_{\rm part} \rangle/2$, plotted as a function of $\langle N_{\rm part} \rangle$ in pO and OO collisions at $\sqrt{s_{\rm NN}}=9.61$ and $7$~TeV, respectively, using the hybrid model used in this work. The presence of $\alpha$-clusters in $^{16}$O yields more charged particles towards high-$N_{\rm part}$, and significantly lower yield in low-$N_{\rm part}$ as compared to the same numbers from the collisions of Woods--Saxon type $^{16}$O nuclei in OO collisions. This is understood as a fact that the $\alpha$-cluster density profile samples the nucleons inside a tight radius, making the nuclear geometry more compact than the Woods--Saxon type sampling, which has an elongated, smooth tail in its probability distribution that can go up to large radial distances. Therefore, more matter is squeezed up at a shorter distance from the center of the nucleus for the $\alpha$-clustered density profile than the Woods--Saxon type profile. This possibly leads to a higher $\langle N_{\rm coll}\rangle$ towards the central collisions (high-$N_{\rm part}$) and explains the observed higher yield than the Woods--Saxon case.

On the other hand, in pO collisions, a slightly different trend is seen. Here,  $\langle dN_{\rm ch}/d\eta \rangle/(\langle N_{\rm part} \rangle/2)$ for $\alpha$-cluster case is higher for the intermediate-$\langle N_{\rm part}\rangle$ (midcentral collisions). This is because, in the $\alpha$-cluster profile, most of the matter is concentrated near the RMS radius of the $\alpha$-particle, while the center is mostly empty, unlike the Woods--Saxon profile, where the nuclear density is higher at the center. Thus, pO collisions near mid-central class lead to a higher $N_{\rm coll}$ for the clustered nuclear profile, while the central pO collisions with clustered structure of oxygen (high-$\langle N_{\rm part}\rangle$) have relatively smaller yield compared to the Woods--Saxon profile. Interestingly, the $\langle dN_{\rm ch}/d\eta \rangle/(\langle N_{\rm part} \rangle/2)$ of OO for both nuclear profiles shows a qualitative sharp rise towards $N_{\rm part}\lesssim 2A$, where $A$ is the mass number of the colliding nuclei. In Ref.~\cite{ALICE:2018cpu}, it is shown that for a specific $N_{\rm part}$, the value of $\langle dN_{\rm ch}/d\eta \rangle/(\langle N_{\rm part} \rangle/2)$ in Xe--Xe is higher as compared to Pb--Pb collisions. This indicates that the number of binary collisions in Xe--Xe is higher than Pb--Pb for similar values of $N_{\rm part}$~\cite{ALICE:2018cpu, ALICE:2015juo}. Similarly, in Fig.~\ref{fig:dNchdeta}, a higher value of $\langle dN_{\rm ch}/d\eta \rangle/(\langle N_{\rm part} \rangle/2)$ in pO collisions than OO near $\langle N_{\rm part}\rangle\simeq6$ for both the nuclear density profiles is also attributed to a higher $\langle N_{\rm coll}\rangle$ in pO collisions as compared to OO collisions. 

\begin{figure*}[ht!]
\centering
\includegraphics[scale=0.44]{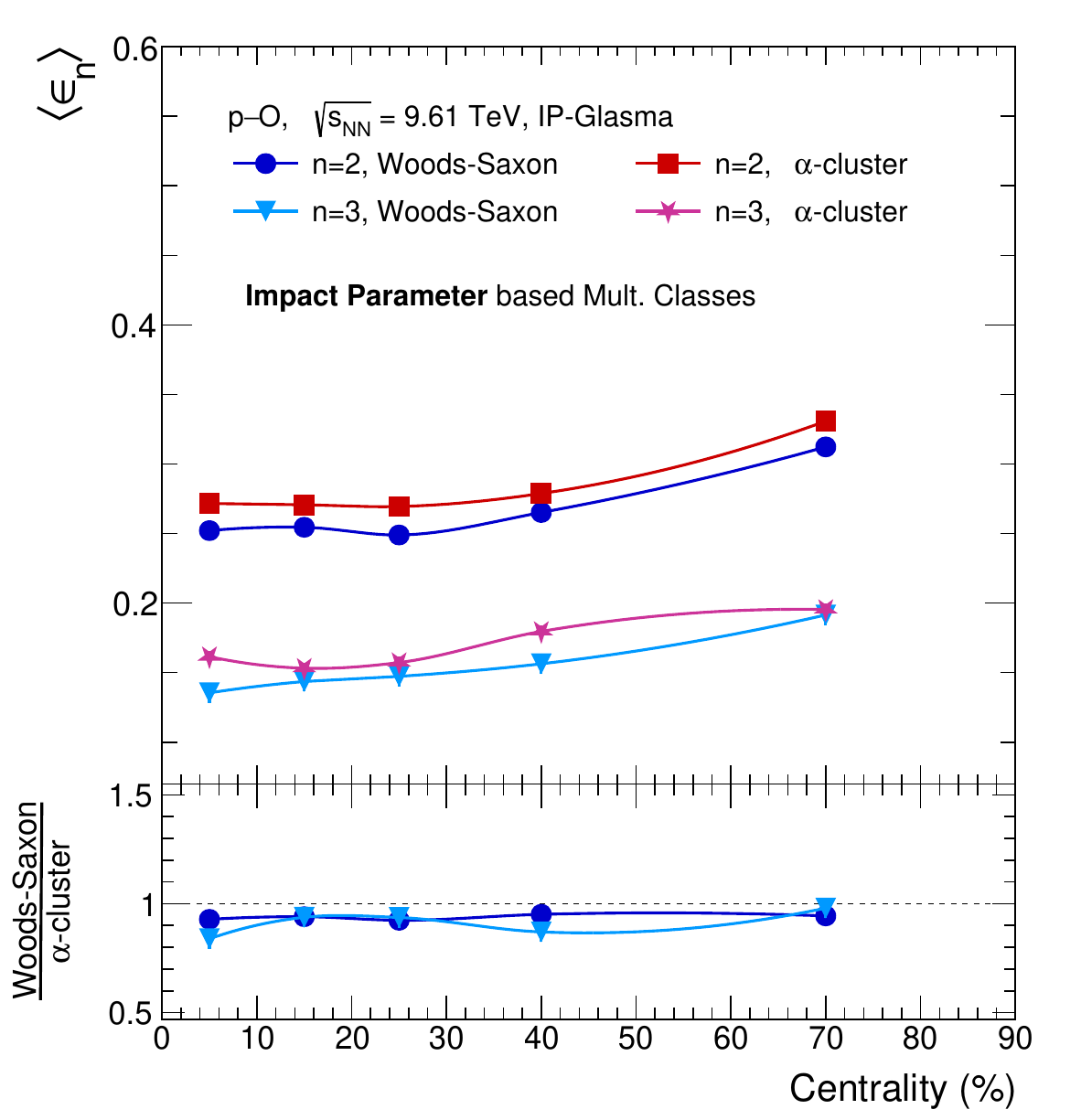}
\includegraphics[scale=0.44]{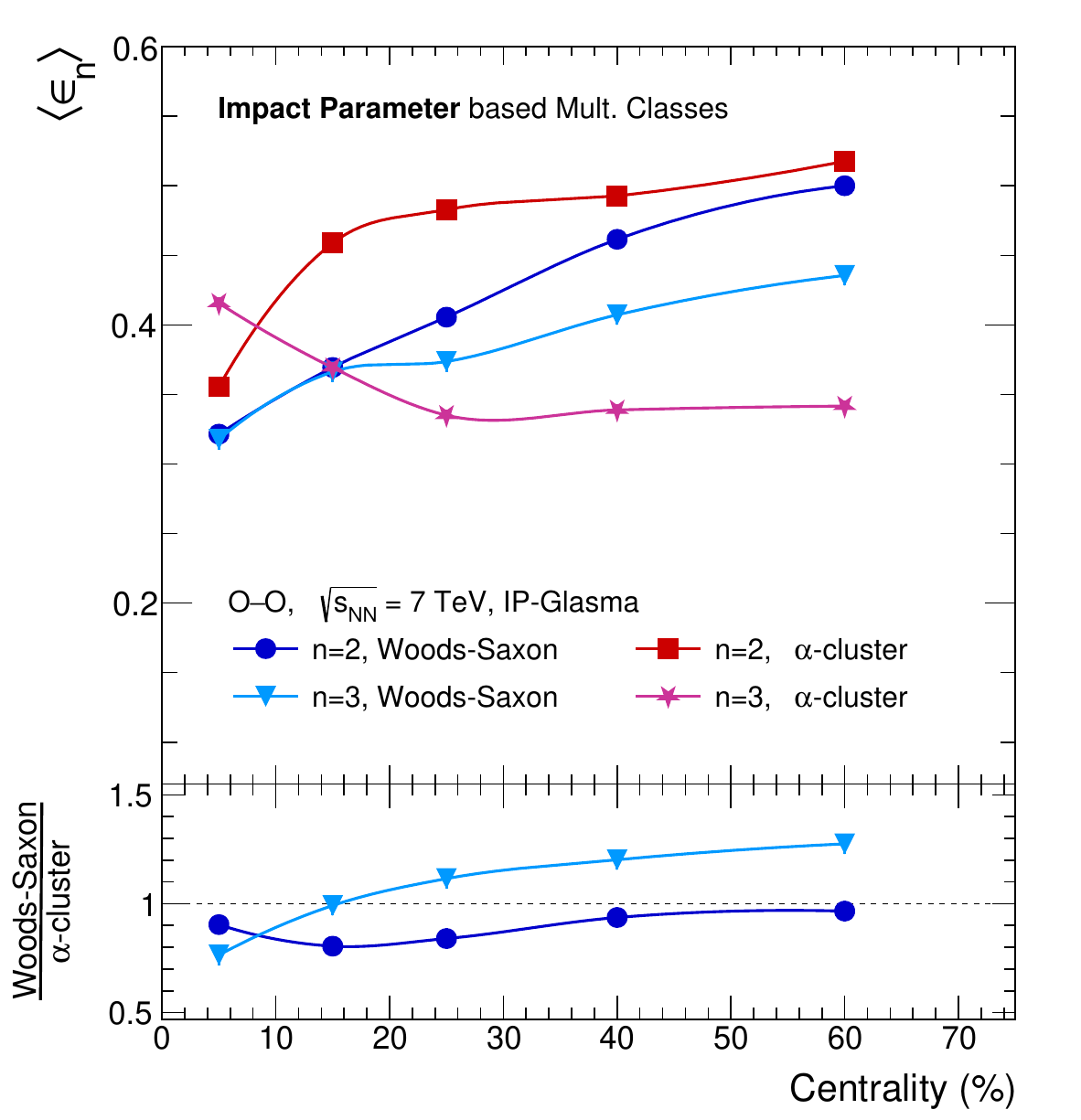}


 \caption{Upper panel shows the average eccentricity and triangularity as a function of impact-parameter based centrality class (\%) for pO collisions at $\sqrt{s_{\rm NN}}=9.61$ TeV (left) and OO collisions at $\sqrt{s_{\rm NN}}=7$ TeV (right) for Woods--Saxon and  $\alpha-$clustered nuclear density profiles, using IP-Glasma. The ratios of $\langle\epsilon_{2}\rangle$ and $\langle\epsilon_{3}\rangle$ from Woods--Saxon to $\alpha-$clustered nuclear density profile are shown in the lower panel. Estimated statistical uncertainties are well within the marker sizes.} 
\label{fig:ecc}
\end{figure*}

\subsection{Initial spatial anisotropy}
\label{sec:resultseccentricity}
Figure~\ref{fig:ecc} shows the collision centrality dependence of IP-Glasma generated initial geometrical eccentricities ($\langle \epsilon_2 \rangle$ and $ \langle \epsilon_3 \rangle$) in pO collisions at $\sqrt{s_{\rm NN}}=9.61$ TeV (left) and OO collisions at $\sqrt{s_{\rm NN}}=7$ TeV (right), with both Woods--Saxon and $\alpha$-clustered type nuclear density for $^{16}$O. The bottom panels show the ratio of eccentricities of the same order between Woods--Saxon and $\alpha$-clustered density profiles. 

In case of pO collisions, $\langle\epsilon_{2}\rangle$ increases from central to peripheral collisions, for both Woods--Saxon and $\alpha$-cluster type density profiles. The value of $\langle\epsilon_{2}\rangle$ towards the peripheral collisions is around $25\%$ higher than the most central pO collisions for both types of nuclear density profiles. Although both the density profiles show similar trends for the centrality dependence of $\langle\epsilon_{2}\rangle$, the values of $\langle\epsilon_{2}\rangle$ from the $\alpha$-cluster type density profile dominate over the Woods--Saxon density profile across all the centrality bins.
A similar centrality dependence is also observed for $\langle\epsilon_{3}\rangle$, with $\alpha$-cluster type density profile dominating over the Woods--Saxon density profile across all the centrality bins, except towards the most peripheral case, where the values of $\langle\epsilon_{3}\rangle$ from both types of density profiles converge. This behavior is also represented in the bottom ratio plot of the left panel, where the ratio is less than one for both $\langle\epsilon_{2}\rangle$ and $\langle\epsilon_{3}\rangle$. Clearly, $\langle\epsilon_{2}\rangle$ is quantitatively greater than $\langle\epsilon_{3}\rangle$ for pO collisions, consistent with the observations from heavy-ion collisions such as p--Pb and Pb--Pb collisions~\cite{ALICE:2019zfl, ALICE:2016ccg}. 

In the right panel of Fig.~\ref{fig:ecc} for OO collisions involving Woods--Saxon density profile, $\langle\epsilon_{2}\rangle$ shows an increasing trend from central to peripheral collisions; however, this increase is steadier than in pO collisions, and it leads to a rise of more than $40\%$ of its value towards the peripheral collisions compared to the most central case. On the other hand, $\langle\epsilon_{2}\rangle$ for the $\alpha$-cluster case dominates over the Woods--Saxon profile from central to mid-central collisions, yet approaches similar values towards the peripheral case. Thus, a noticeable nuclear density profile dependence of $\langle\epsilon_{2}\rangle$ is observed, with the eccentricity in the 0--70\% centrality class being clearly higher for the $\alpha$-cluster profile than that for the Woods--Saxon nuclear density profile~\cite{Prasad:2024ahm}. Further, $\langle\epsilon_{3}\rangle$ for OO collisions increases from central to peripheral collisions for the Woods--Saxon profile as in pO collisions, while for the case of $\alpha$-cluster profile, it shows a decreasing-and-saturating trend with the centrality. In the most central OO collisions, $\langle\epsilon_{3}\rangle$ from $\alpha$-clustered nuclear density profile is observed to be $\approx$ $30\%$ higher than that of the Woods--Saxon case. This feature is also visible in the AMPT results for OO collisions at $\sqrt{s_{\rm NN}}$ = 7 TeV~\cite{Behera:2023nwj}, and should arise as a consequence of the overlap of two triangular sides of the colliding tetrahedral oxygen nuclei with the $\alpha$-clustered nuclear density profile. The occurrence of this unique behaviour in $\alpha$-clustered OO collisions is further confirmed by studying the centrality dependence of $\langle\epsilon_{3}\rangle$ with varying $\alpha$-cluster parameters, where the trend of the results was found to be consistent with each other (See Appendix~\ref{AppendixA}). In addition, the value of $\langle\epsilon_{3}\rangle$ from Woods--Saxon dominates over the $\alpha$-clustered profile in the off-central OO collisions, and this can be quantitatively analyzed from the bottom panel of the right plot of Fig.~\ref{fig:ecc}.

\subsection{Anisotropic flow coefficients}
\label{sec:vnvscent}

\begin{figure*}[ht!]
\centering
\includegraphics[scale=0.44]{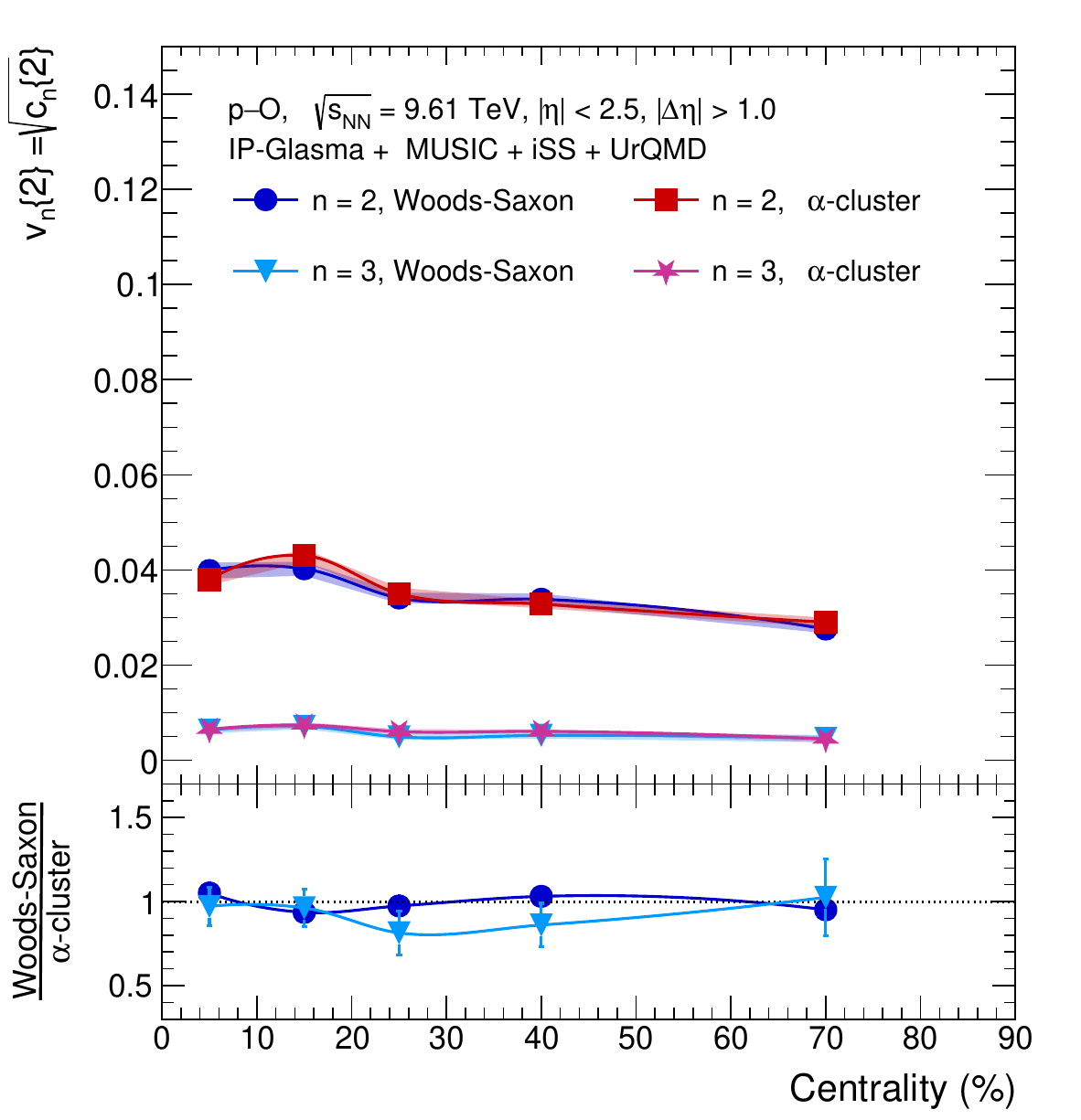}
\includegraphics[scale=0.44]{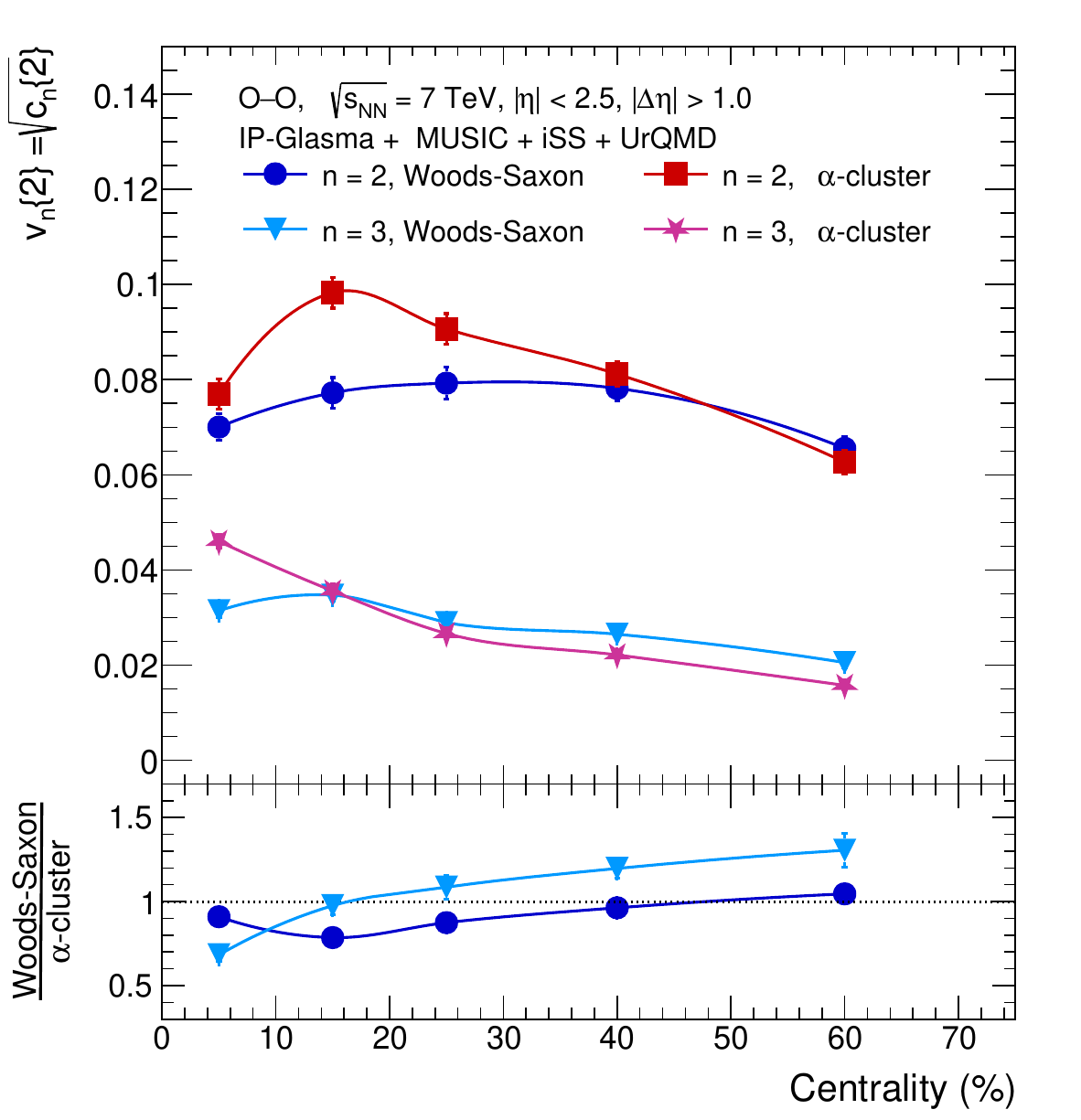}

\caption{Elliptic and triangular flow calculated using the two-particle Q-cumulant method as a function of collision centrality for pO collisions at $\sqrt{s_{\rm NN}}=9.61$ TeV (left) and OO collisions at $\sqrt{s_{\rm NN}}=7$ TeV (right) from the IP-Glasma + MUSIC + iSS + UrQMD framework. Results include both Woods--Saxon and $\alpha-$clustered type $^{16}\rm O$ nuclei. The ratio of $v_{\rm 2}\{2, |\Delta\eta|>1.0\}$ and $v_{\rm 3}\{2, |\Delta\eta|>1.0\}$ from Woods--Saxon to $\alpha-$cluster density profile is shown in the lower panels.}
\label{fig:vn}
\end{figure*}

Using the two-particle Q-cumulant method, the calculation for elliptic and triangular flow is performed for all charged particles in $|\eta|<$ 2.5, in pO and OO collisions at $\sqrt{s_{\rm NN}}=9.61$ TeV and $\sqrt{s_{\rm NN}}=7$ TeV, respectively. A pseudorapidity gap of $|\Delta\eta| > 1.0$ is imposed between the particle pairs to remove short-range jet-like correlations, if any. In addition to the pseudorapidity gap, $|\Delta\eta|>1$, estimation with $|\Delta\eta|>2$ \textit{i.e.} $v_{n}\{2, |\Delta\eta|> 2\}$ was also performed, so as to ensure maximum reduction of non-flow contributions. Since the results showed little or no difference from the estimated $v_{n}\{2, |\Delta\eta|> 1\}$, we resort to estimation with $|\Delta\eta|>1$.

In Fig.~\ref{fig:vn}, both $v_{2}\{2,|\Delta\eta| > 1.0\}$ and $v_{3}\{2,|\Delta\eta| > 1.0\}$ as a function of collision centrality are shown, for both Woods--Saxon and $\alpha$-cluster type $^{16}\rm O$ nuclei in pO collisions at $\sqrt{s_{\rm NN}}=9.61$~TeV and OO collisions at $\sqrt{s_{\rm NN}}=7$~TeV, from the IP-Glasma + MUSIC + iSS + UrQMD framework. As one can see from the left plot of Fig.~\ref{fig:vn}, the nuclear density profile dependence for $v_{2}\{2,|\Delta\eta| > 1.0\}$ and $v_{3}\{2,|\Delta\eta| > 1.0\}$ is almost negligible in pO collisions. For both nuclear density profiles, $v_{2}\{2,|\Delta\eta| > 1.0\}$ drops gradually from the most central to peripheral pO collisions, with the $\alpha$-cluster profile showing slightly higher values than the Woods--Saxon profile in the 10--20\% centrality class, appearing as a small bump-like structure. On the other hand, $v_{3}\{2,|\Delta\eta| > 1.0\}$ shows no significant centrality dependence, and the trends for both Woods--Saxon and $\alpha$-cluster profile cases almost overlap with one another. 

On the contrary, for OO collisions, both $v_{2}\{2,|\Delta\eta| > 1.0\}$ and $v_{3}\{2,|\Delta\eta| > 1.0\}$ have a strong dependence on the choice of nuclear density profiles. $v_{2}\{2,|\Delta\eta| > 1.0\}$ for the Woods--Saxon case has a smooth increase from central to midcentral collisions and then a relatively gradual decrease towards the peripheral collisions. For the OO collisions with $\alpha$-clustered profile for $^{16}\rm O$ nuclei, the initial rise until midcentral collisions and the following drop towards peripheral collisions are much steeper compared to the Woods--Saxon case. In fact, it is seen in the $\alpha$-cluster case that the $v_{2}\{2,|\Delta\eta| > 1.0\}$ attains a peak at 10--20\% centrality class ($\approx$ 30\% higher than the corresponding $v_{2}$ from Woods--Saxon profile). This is also similar to the small bump observed in the case of pO collisions for the same centrality of collisions. However, as one moves from mid-central to peripheral OO collisions, the magnitude of $v_{2}\{2,|\Delta\eta| > 1.0\}$ steeply falls to almost half of its peak value, with $v_{2}\{2,|\Delta\eta| > 1.0\}$ from Woods--Saxon case starting to dominate over the elliptic flow from the $\alpha$-cluster profile case.

In addition, $v_{3}\{2,|\Delta\eta| > 1.0\}$ has a smooth decrease from the central to peripheral OO collisions for both types of nuclear density profiles studied in this work\footnote{A similar trend is observed in the experimental measurements of $v_3$ versus centrality reported by ALICE and ATLAS collaborations~\cite{ALICE:2025luc, ATLAS:2025nnt}.}; the $\alpha$-cluster case has a higher value of $v_{3}\{2,|\Delta\eta| > 1.0\}$ for the most central collisions compared to the same from the Woods--Saxon case. However, for the rest of the centrality bins, the Woods--Saxon type profile leads the value of $v_{3}\{2,|\Delta\eta| > 1.0\}$. 

The major contribution to the nuclear density profile dependence of $v_{n}$ arises from the initial state eccentricities.  For instance, $v_{2}$ in the 10--20\% centrality and $v_{3}$ in the 0--10\% centrality class for the $\alpha$-clustered profile in OO collisions are reflections of the corresponding $\langle\epsilon_{2}\rangle$ and $\langle\epsilon_{3}\rangle$. However, the effect of the centrality dependence of $\langle\epsilon_{2}\rangle$ and $\langle\epsilon_{3}\rangle$ is not very well carried forward to the final state, beyond mid-central collisions, owing to the smaller size and lifetime of the system formed. However, for the case of OO collisions, the differences in the initial spatial anisotropy between the Woods--Saxon and $\alpha$-cluster profile propagate nicely to the final state azimuthal anisotropy, as visible from the bottom ratio panels of Fig.~\ref{fig:ecc} and Fig.~\ref{fig:vn}. A simple cross-check on the linear variation of $\langle v_{2} \rangle$ with respect to $\epsilon_{2}$ for different classes of collision centrality for all four collision systems is presented in Fig.~\ref{fig:e2v2LINEAR} in Appendix~\ref{AppendixB}.

In a nutshell, the distinction between the $\alpha$-clustered and Woods--Saxon nuclear density profiles in pO collisions is not as prominent as it is for the OO collisions in the measurements of anisotropic flow coefficients, \textit{i.e.}, $v_2$ and $v_3$. These differences may arise due to the difference in system size, as pO collisions result in a smaller system compared to OO collisions. In proton–nucleus collisions, the proton-sized projectile typically interacts with a limited, local region of the nucleus and therefore samples mainly the local fluctuations rather than the full geometric overlap. In contrast, in a scenario of many nucleon–nucleon encounters across the overlap region, for instance, in OO collisions, each nucleon in each of the oxygen nuclei acts as a small probe, together reflecting the geometric structure of the system. As a result, the imaging power could be understood to be significantly greater in OO than in pO collisions. This makes OO a more natural “bridge” system between pp and Pb–-Pb when the goal is to study how global geometry affects collective observables.

\subsection{Multiplicity dependence of $v_{n}$ and $v_{n}$/$\langle\epsilon_{n}\rangle$}

\begin{figure*}[ht!]
\centering
\includegraphics[scale=0.44]{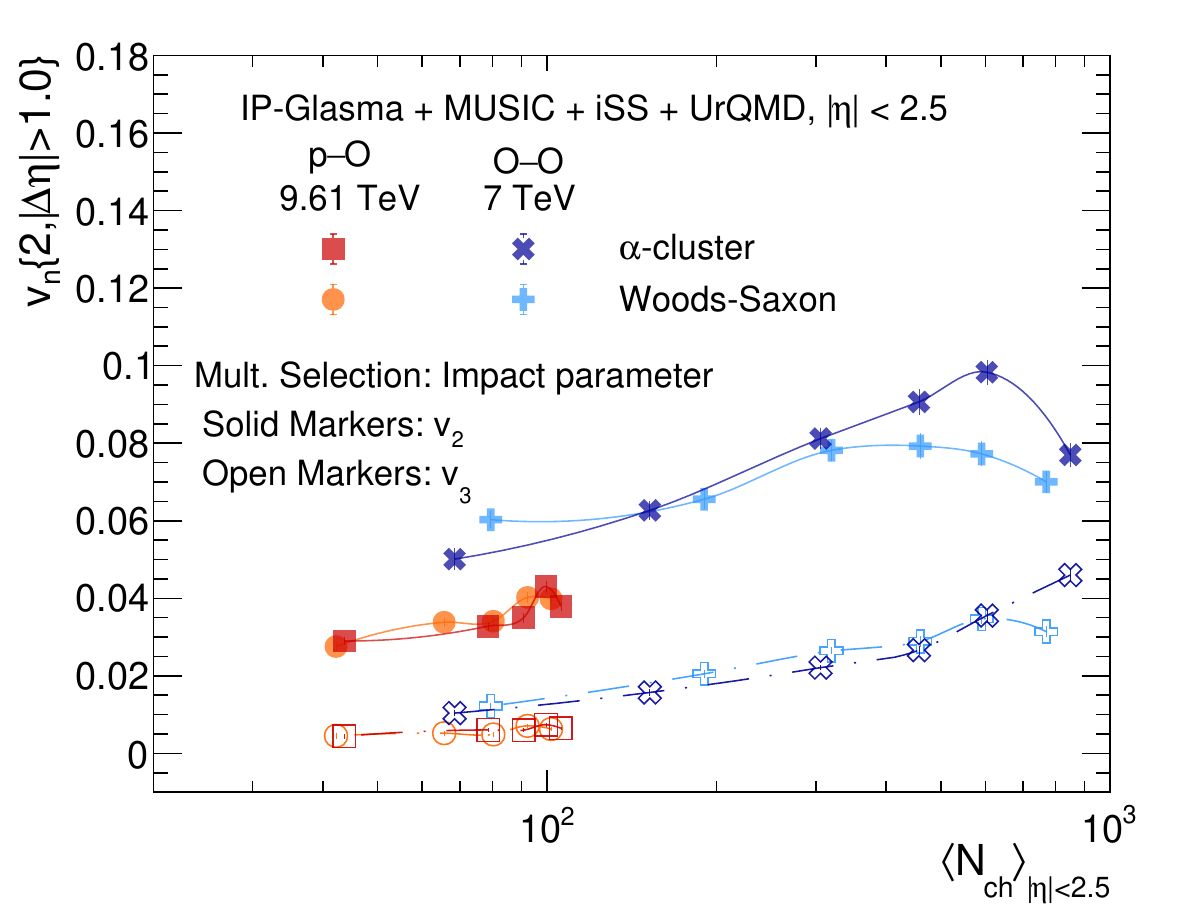}
\includegraphics[scale=0.44]{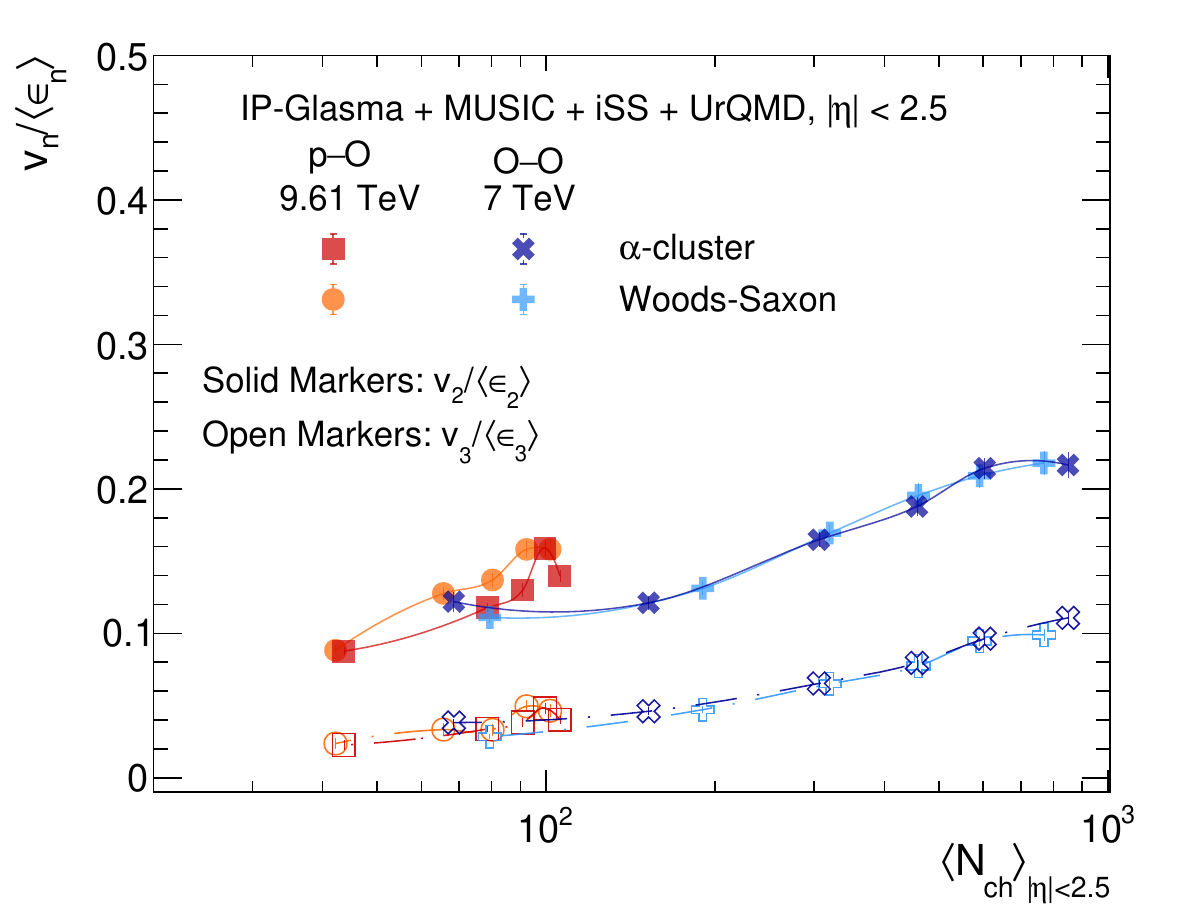}
\caption{$v_{n}\{2, |\Delta\eta|>1.0\}$ (left panel) and $v_{n}\{2, |\Delta\eta|>1.0\}$/$\langle\epsilon_{n}\rangle$ (right panel) as a function of average charged-particle multiplicity ($\langle N_{\rm ch}\rangle_{|\eta|<2.5}$) measured in $|\eta|<2.5$ for pO collisions at $\sqrt{s_{\rm NN}}=9.61$ TeV and OO collisions at $\sqrt{s_{\rm NN}}=7$ TeV using IP-Glasma + MUSIC + iSS + UrQMD model, for Woods--Saxon and $\alpha-$cluster density profiles of $^{16}\rm O$ nuclei. Estimated statistical uncertainties are well within the marker sizes.}
\label{fig:vn_<Nch>}
\end{figure*}

One of the major motivations for the recently performed pO and OO collisions at the LHC comes from the fact that these collision systems form a perfect system size to fill the multiplicity gap between the collision systems, pp, p--Pb, and Pb--Pb. Or in other words, studying pO and OO collisions (where produced multiplicity is expected to be intermediate to that of other systems mentioned above) helps us in understanding how the signatures of hydrodynamic behaviour (or any other signatures of QGP) undergo a transition as one moves down from high to low multiplicity, analogously, large to small systems~\cite{Brewer:2021kiv}. It is already quite well known that the particle multiplicity plays a crucial role in transforming the initial eccentricities to the final-state flow. Compared to a low multiplicity event, a collision achieving a larger multiplicity would naturally imply that the hydrodynamic phase during its evolution is longer, which results in an efficient (complete) transformation of initial spatial anisotropies to final-state momentum anisotropies. As observed in Fig.~\ref{fig:vn}, OO collisions show significant dependence of $v_{n}$ on the initial clustering, and the initial-state effects are reflected better in OO than in pO collisions. In this context, studying $\langle\epsilon_{n}\rangle$ and $v_{n}$ for their charged particle multiplicity dependence would help us discern where the system behaviour changes, in turn enabling us to quantify the hydrodynamic limit.

Shown in the left plot of Fig.~\ref{fig:vn_<Nch>} is the elliptic and triangular flow as a function of average charged-particle multiplicity in the pseudorapidity range $|\eta|<2.5$ and $p_{\rm T} > 0.3$ GeV/$c$ ($\langle N_{\rm ch}\rangle_{|\eta|<2.5}$), for pO collisions at $\sqrt{s_{\rm NN}}=9.61$ TeV and OO collisions at $\sqrt{s_{\rm NN}}=7$ TeV, generated using IP-Glasma + MUSIC + iSS + UrQMD framework. Since the $\langle N_{\rm ch}\rangle$ is calculated based on the impact parameter-based event classes, Fig.~\ref{fig:vn_<Nch>} is visually the mirror image of Fig.~\ref{fig:vn}. However, the purpose of Fig.~\ref{fig:vn_<Nch>} is to understand the multiplicity dependence of $v_{2}\{2,|\Delta\eta| > 1.0\}$ and $v_{3}\{2,|\Delta\eta| > 1.0\}$ for two different collision systems \textit{i.e.} pO and OO, for both $\alpha$-clustered and Woods--Saxon nuclear density profiles of colliding $^{16}\rm O$, to be in line with the experimental results which are presented as a function of multiplicity. 

As can be seen, $v_{2}\{2,|\Delta\eta| > 1.0\}$ has the maximum value at $\langle N_{\rm ch} \rangle \simeq$ 600 for OO collisions involving $\alpha$-clustered nuclear density profile, while this sharp peak is not observed in the case of the Woods--Saxon profile for OO, as already discussed earlier. If compared with the results in Ref.~\cite{ALICE:2019zfl},  $v_{2}\{2,|\Delta\eta| > 1.0\}$ achieved by the 10-20\% central OO collisions involving compact $\alpha$-clusters, is comparable to the elliptic flow attained in mid-central Pb--Pb collisions. Now, as the particle multiplicity in the desired pseudorapidity ($\eta$) range falls by an order for OO collisions ($\langle N_{\rm ch} \rangle \simeq$ 75), the $v_{2}\{2,|\Delta\eta| > 1.0\}$ is reduced by 40\%. Interestingly, even in the highest multiplicity pO collisions with $\alpha$-clustered $^{16}\rm O$ nuclei (where $\langle N_{\rm ch} \rangle \simeq$ 100), we see a tiny bump like structure in the trend for $v_{2}\{2,|\Delta\eta| > 1.0\}$, similar to the observation in OO collisions with $\alpha$-cluster profiles. Additionally, in the case of the Woods--Saxon profile, the observation of a gradual decrease of $v_{2}\{2,|\Delta\eta| > 1.0\}$ from high to low multiplicity classes, with no abrupt jumps, is common for both OO and pO collisions. In the case of triangular flow, $v_{3}\{2,|\Delta\eta| > 1.0\}$ obtained in high to intermediate multiplicity pO collisions is very similar to that obtained in the lowest multiplicity OO collisions, which is expected from Ref.~\cite{ALICE:2019zfl}. Though a moderate dependence is seen on nuclear density profiles for $v_{3}\{2,|\Delta\eta| > 1.0\}$ in OO collisions, there is no significant influence of $\alpha$-clustering on $v_{3}\{2,|\Delta\eta| > 1.0\}$ in the case of pO collisions. 

Scaling $v_{n}$ by $\langle\epsilon_{n}\rangle$ is a quantified way to understand the extent to which the initial spatial anisotropy is converted to the final state momentum anisotropy and hence is believed to probe the transport properties of the medium, and the applicability of hydrodynamic models to the system under study. In the right plot of Fig.~\ref{fig:vn_<Nch>}, thus $v_{2}$/$\langle\epsilon_{2}\rangle$ and $v_{3}$/$\langle\epsilon_{3}\rangle$ are presented as a function of $\langle N_{\rm ch}\rangle_{|\eta|<2.5}$ for pO collisions at $\sqrt{s_{\rm NN}}=9.61$ TeV and OO collisions at $\sqrt{s_{\rm NN}}=7$ TeV, generated using IP-Glasma + MUSIC + iSS + UrQMD framework, for Woods--Saxon and $\alpha$-cluster density profiles of $^{16}\rm O$ nucleus. The scaled ratios now reveal an interesting trend: the increase of $v_{3}$/$\langle\epsilon_{3}\rangle$ with $\langle N_{\rm ch}\rangle_{|\eta|<2.5}$ experiences a smooth transition from pO to OO collisions where the rate of increment seems to be similar for both collision systems. Similarly, $v_{2}$/$\langle\epsilon_{2}\rangle$ also increases with $\langle N_{\rm ch}\rangle_{|\eta|<2.5}$; the slight bump-like structure observed in the trends of $v_{2}$ in pO collisions still persists in the ratio $v_{2}$/$\langle\epsilon_{2}\rangle$, while it is smoothened out in the case of OO collisions. However, the nuclear density profile dependence is no longer seen to remain once scaling of $v_{n}$ with $\langle\epsilon_{n}\rangle$ is performed, except for the $v_{2}$/$\langle\epsilon_{2}\rangle$ case of pO collisions. This is a testimony that, in this work, the choice of nuclear density profile has little to no effect on the medium formed.

\subsection{Impact parameter dependence of $v_n$}

\begin{figure}[ht!]
\centering
\includegraphics[scale=0.44]{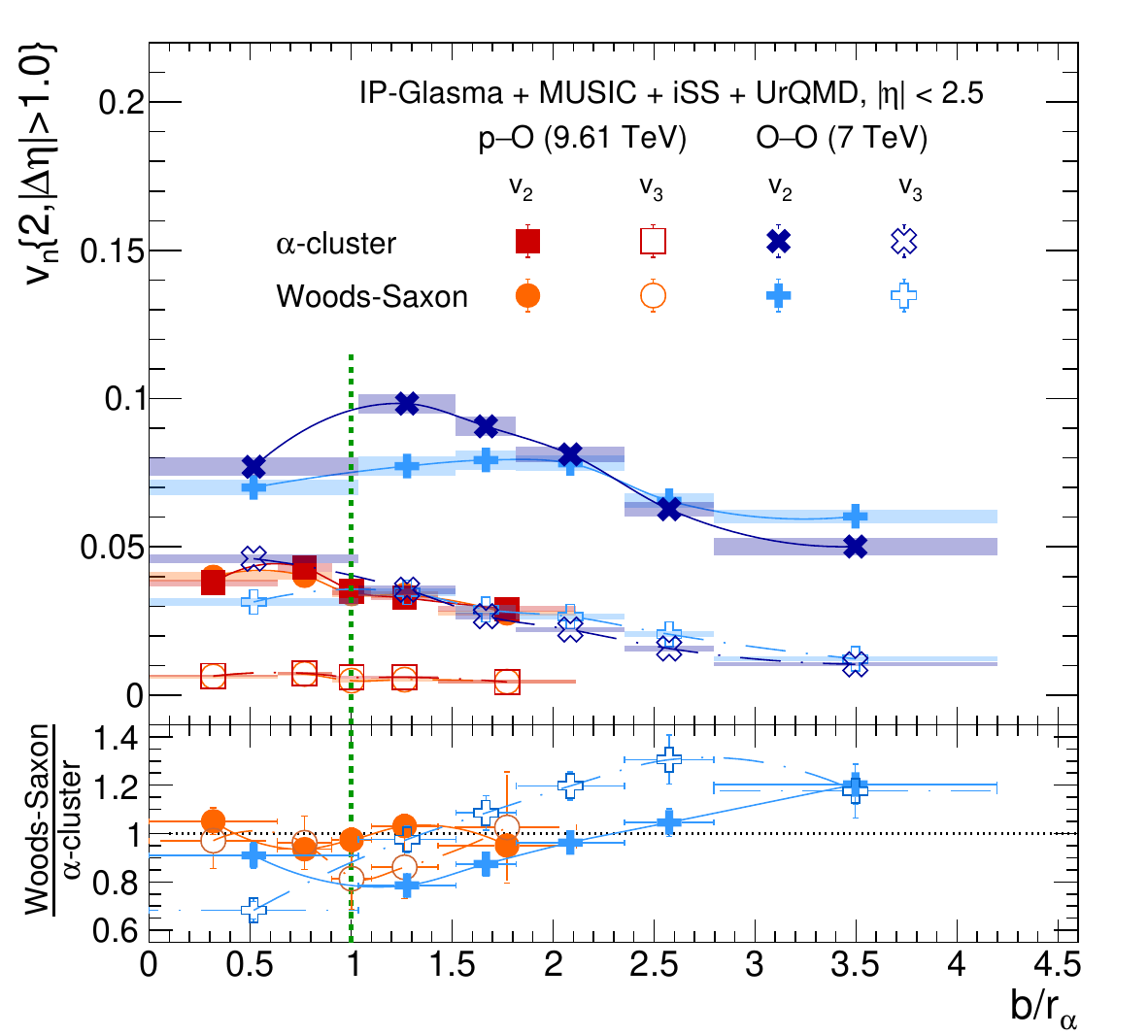}
\caption{$v_{\rm 2}\{2, |\Delta\eta|>1.0\}$ and $v_{\rm 3}\{2, |\Delta\eta|>1.0\}$ as a function of $\rm b/\rm r_{\alpha}$ 
 (where, $\rm r_{\alpha}=1.676$ fm) for pO collisions at $\sqrt{s_{\rm NN}}=9.61$ TeV and OO collisions at $\sqrt{s_{\rm NN}}=7$ TeV using IP-Glasma + MUSIC + iSS + UrQMD model, for Woods--Saxon and $\alpha-$cluster density profiles of colliding $^{16}\rm O$ nucleus. Error bands incorporate the estimated statistical uncertainties.}
\label{fig:vn_bralpha}
\end{figure}



In Figs.~\ref{fig:vn} and \ref{fig:vn_<Nch>}, it was observed that the effect of clustered structure is mostly highlighted in the high-multiplicity or most central regions. Thus, it is important to quantify the regions as a function of impact parameter ($b$) that is capable of highlighting the effects in the final state flow coefficients for the choice of nuclear profile ($\alpha$-clustered versus Woods--Saxon). This would help us understand whether the size of the $\alpha$-particle (radius of $\alpha$-particle, $\rm r_{\alpha}=1.676$ fm) has any role to play. In Fig.~\ref{fig:vn_bralpha}, we show $v_{n}\{2, |\Delta\eta|>1.0\}$ as a function of impact parameter scaled with the RMS radius of $\alpha$-particle ($^{4}\text{He}$ nucleus) ($b/r_{\alpha}$) for both pO and OO collisions at $\sqrt{s_{\rm NN}}=9.61$ TeV and $7$ TeV, respectively, using IP-Glasma + MUSIC + iSS + UrQMD. Figure~\ref{fig:vn_bralpha} would aid in quantitatively conceiving the effect of the clustered structure in the final state flow observables as a function of $b/r_{\alpha}$. Interestingly, in OO collisions, one finds that the dependence of elliptic and triangular flow on nuclear density profile is maximized near $b/r_{\alpha}\simeq 1$. Here, $v_{2}\{2, |\Delta\eta|>1.0\}$ near $b=r_{\alpha}$ shows a peak like structure for the $\alpha$-cluster case and is absent for the Woods--Saxon profile, the effect of which originates from the initial eccentricity, as shown in Fig.~\ref{fig:ecc}. Similarly, a higher $v_{3}\{2, |\Delta\eta|>1.0\}$ for $b/r_{\alpha}<1$ is also attributed to the $\alpha$-clustered case which is absent for the Woods--Saxon profile.
Additionally, in Fig.~\ref{fig:vn_bralpha} it can be observed that, in OO collisions, towards the higher impact parameter values, the Woods--Saxon nuclear profile has larger value of $v_{2}\{2, |\Delta\eta|>1.0\}$ and $v_{3}\{2, |\Delta\eta|>1.0\}$ as compared to the $\alpha$-clustered structure, also reflected in the bottom panel where the ratio of Woods--Saxon to $\alpha$-clustered structure is larger than one. This is similar to the observations made in the right panel of Fig.~\ref{fig:vn} and~\ref{fig:vn_<Nch>}, where $v_{2}\{2, |\Delta\eta|>1.0\}$ and $v_{3}\{2, |\Delta\eta|>1.0\}$ in mid-central and peripheral OO collisions deviate significantly from that of $\langle\epsilon_2\rangle$ and $\langle\epsilon_3\rangle$, respectively, in Fig.~\ref{fig:ecc}.
This could be attributed to the combined effects of the smaller lifetime of the fireball towards the mid-central or peripheral collisions and the difference in the medium response to the evolution of $v_2$ from $v_3$. Additional contributions could arrive from the large fluctuations of anisotropic flow coefficients due to a smaller number of charged particles.
However, due to small particle multiplicity in pO collisions, no such explicit effects of clustered nuclear structure are observed throughout $b/r_{\alpha}$ regions. The variations of $v_{n}\{2, |\Delta\eta|>1.0\}$ with the choice of nuclear density profile are well reflected in the bottom ratio panel.
Therefore, in brief from Fig.~\ref{fig:vn_bralpha}, it is clear that the effects of clustered density profile are well observed in the region $b/r_{\alpha}\lesssim 1$. The sought-after effects are not visible when the particle multiplicity is small, like in the pO collisions.

To have an extended understanding of this pronounced effect at $b = r_{\alpha}$, a comparison of $v_2$ and $v_3$ in OO collisions with different clustered nuclear configurations of $^{16} \rm O$ was performed (See Appendix~\ref{AppendixA}). The results show that a smaller value of $r_{\alpha}$ or a larger value of sidelength leads to larger $v_2$ and $v_3$ for similar values of $b/r_{\alpha}$, where a shift in the peak position of $v_2$ is observed. This enhancement in the values of $v_{n}$ with a decrement in $r_{\alpha}$/$l_{\rm sidelength}$ is consistent with the results observed for OO collisions at $\sqrt{s_{\rm NN}}=  6.5$~TeV~\cite{YuanyuanWang:2024sgp}.
This may indicate that the effects near $b/r_{\alpha}=1$ manifest themselves only in the collisions involving the originally stable $\alpha$-clustered nuclei with very compact structure, which leaves little spatial void. However, a detailed analysis is necessary to establish the empirical description of such effects of initial clustered configuration on final state anisotropic flow coefficients. Though beyond the scope of this specific study, exploring and validating these $r_{\alpha}$-dependent effects in contrast to unclustered Woods--Saxon nuclear profiles in collisions involving $\alpha$-clustered nuclei such as $^{12} \rm C$, $^{20}\rm Ne$ etc. would be immensely beneficial to counter-check our results and if confirmed, to provide a method to estimate the radius of the $\rm ^{4}He$ nucleus.

\section{SUMMARY}
\label{sec:summary}
In this study, the effects of the clustered nuclear structure of $^{16} \rm O$ nuclei are investigated through pO and OO collisions at the LHC energies using IP-Glasma + MUSIC + iSS + UrQMD. The final state anisotropic flow coefficients are studied and are compared with the corresponding spatial anisotropies of the initial collision overlap geometry. Our findings are summarized as follows: 
\begin{enumerate}

\item For both pO and OO collision systems, it is observed that the $\langle\epsilon_{2}\rangle$ rises from central to peripheral collisions with a clear dominance of $\alpha$-cluster over the Woods--Saxon profile case. Similar is the trend for $\langle\epsilon_{3}\rangle$ except for OO collisions with $\alpha$-clustering. Here, the $\langle\epsilon_{3}\rangle$ is found to be maximum in the 0--10\% centrality class and falls thereafter in mid-central collisions. This trend is found to be consistent even in cases where the $\alpha$-clustering parameters are varied.

\item The centrality dependence of initial spatial anisotropies ($\langle\epsilon_{n}\rangle$) is not effectively converted to final-state elliptic and triangular flow, especially in the mid-central to peripheral pO and OO collisions, owing to the less dense systems formed in the off-central collisions. 

\item The nuclear density profile seems to have little influence on $v_{2}$ and $v_{3}$ in pO collisions, while it has a noticeable effect in the case of OO collisions. In spite of this, $v_{2}\{2,|\Delta\eta| > 1.0\}$ is found to achieve its maximum in the 10--20\% centrality class, for both pO and OO collisions involving $\alpha$-clustered $^{16}\rm O$. 

\item Once $v_{n}$ is scaled with the corresponding $\langle\epsilon_{n}\rangle$, the nuclear profile dependence vanishes completely for OO collisions. This scaling also brings the multiplicity-dependent curve of $v_{n}$/$\langle \epsilon_{n}\rangle$ for pO and OO collisions to a single continuous line.

\item In OO collisions, if oxygen nuclei have a compactly arranged $\alpha$-clustered structure, the effects due to $\alpha$-clustering are observed to manifest well in the region $b/r_{\rm \alpha}\lesssim 1$, therefore being comparable with the size of the $^{4}\rm He$.
\end{enumerate}

In this work, we present a systematic study of how the $\alpha$-clustered nuclear structure of the colliding $^{16}\rm O$ nucleus in collision systems ranging from pO to OO collisions impacts the final-state medium anisotropy. This study can further be useful as a hydrodynamic prediction for future experimental investigations of pO and OO collisions at the LHC. A comparison with experimental data would help constrain the nuclear density profile for the oxygen nuclei and aid in tuning other models. Additionally, the present study also provides an interesting prospect to explore the $v_n$ distributions in small systems, through the estimation of event-by-event $v_n$ values, which is not possible in the current framework employed in this paper. The study of $v_n$ and their distributions for different centrality, event-shape, and $\epsilon_n$ bins would enable us to understand their evolution, fluctuations, higher moments, and interplay among different harmonics~\cite{Prasad:2025ezg}, which would in turn enable the heavy-ion physics community to have a better understanding of collectivity in small systems.

\section*{Acknowledgement}
AMKR acknowledges the doctoral fellowships from the DST INSPIRE program of the Government of India. SP acknowledges the doctoral fellowship from the UGC, Government of India. RS sincerely acknowledges the DAE-DST, Government of India, funding under the mega-science project – “Indian participation in the ALICE experiment at CERN” bearing Project No. SR/MF/PS-02/2021-IITI (E-37123). NM is supported by the Academy of Finland through the Center of Excellence in Quark Matter with Grant No. 346328. GGB gratefully acknowledges the Hungarian National Research, Development and Innovation Office (NKFIH) under Contract No. NKFIH NEMZ\_KI-2022-00058, 2024-1.2.5-TET-2024-00022, and Wigner Scientific Computing Laboratory (WSCLAB, the former Wigner GPU Laboratory). The authors gratefully acknowledge the MoU between IIT Indore and Wigner Research Centre for Physics (WRCP), Hungary, for the techno-scientific international cooperation. 

\appendix
\section*{Appendix}

\subsection{Varying $\alpha$-cluster parameters}
\label{AppendixA}

\begin{figure*}[ht!]
\centering
\includegraphics[scale=0.29]{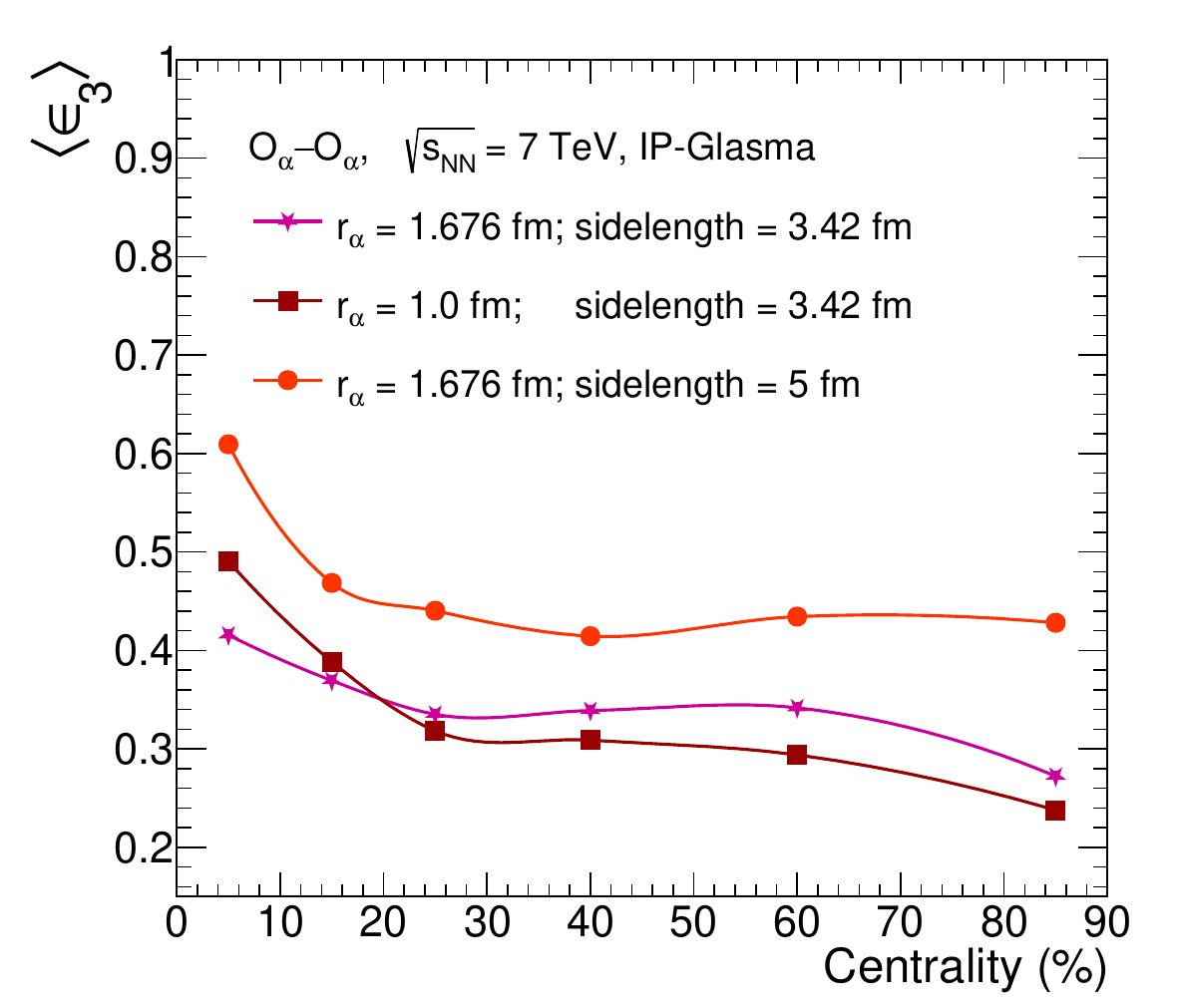}
\includegraphics[scale=0.29]{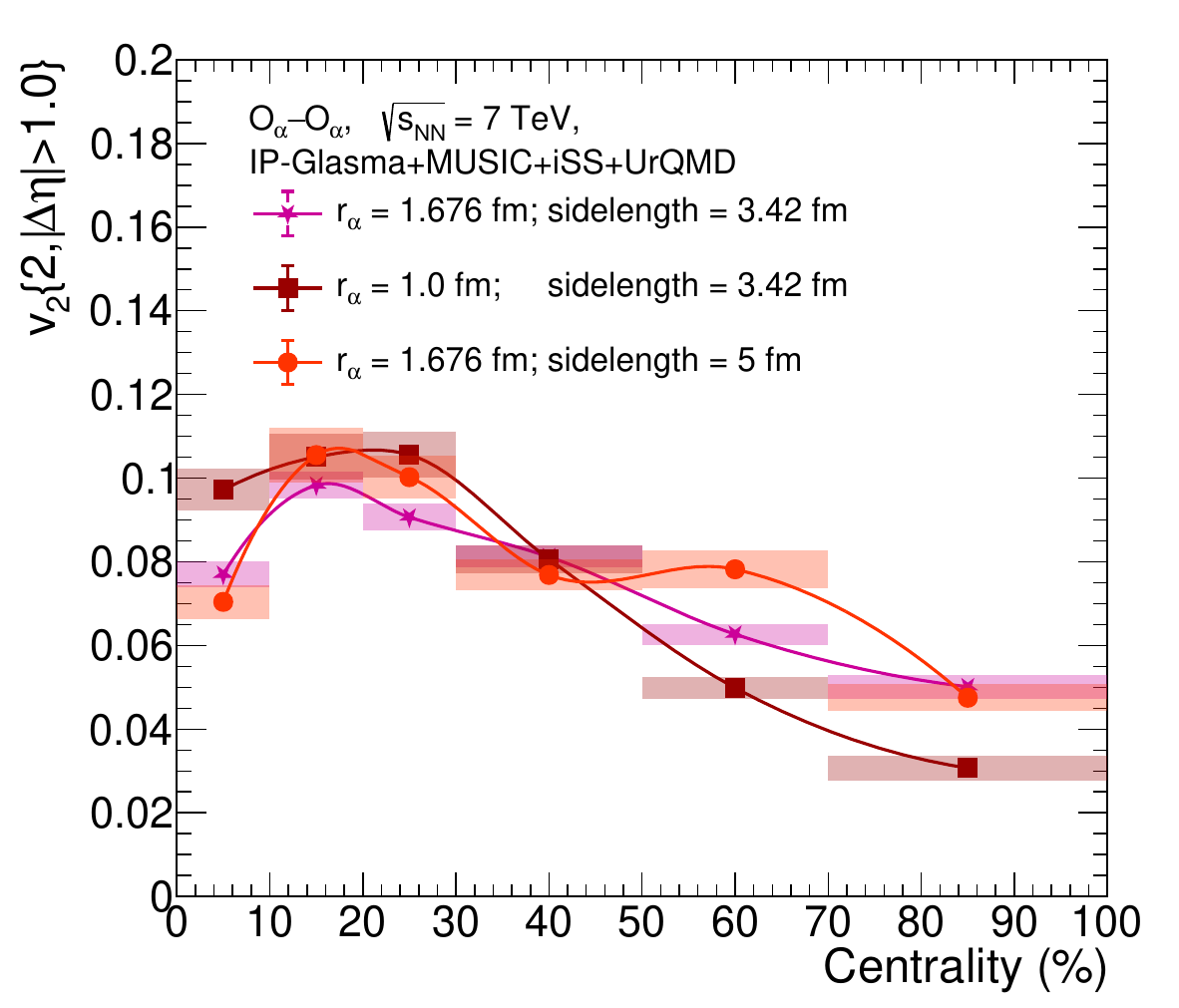}
\includegraphics[scale=0.29]{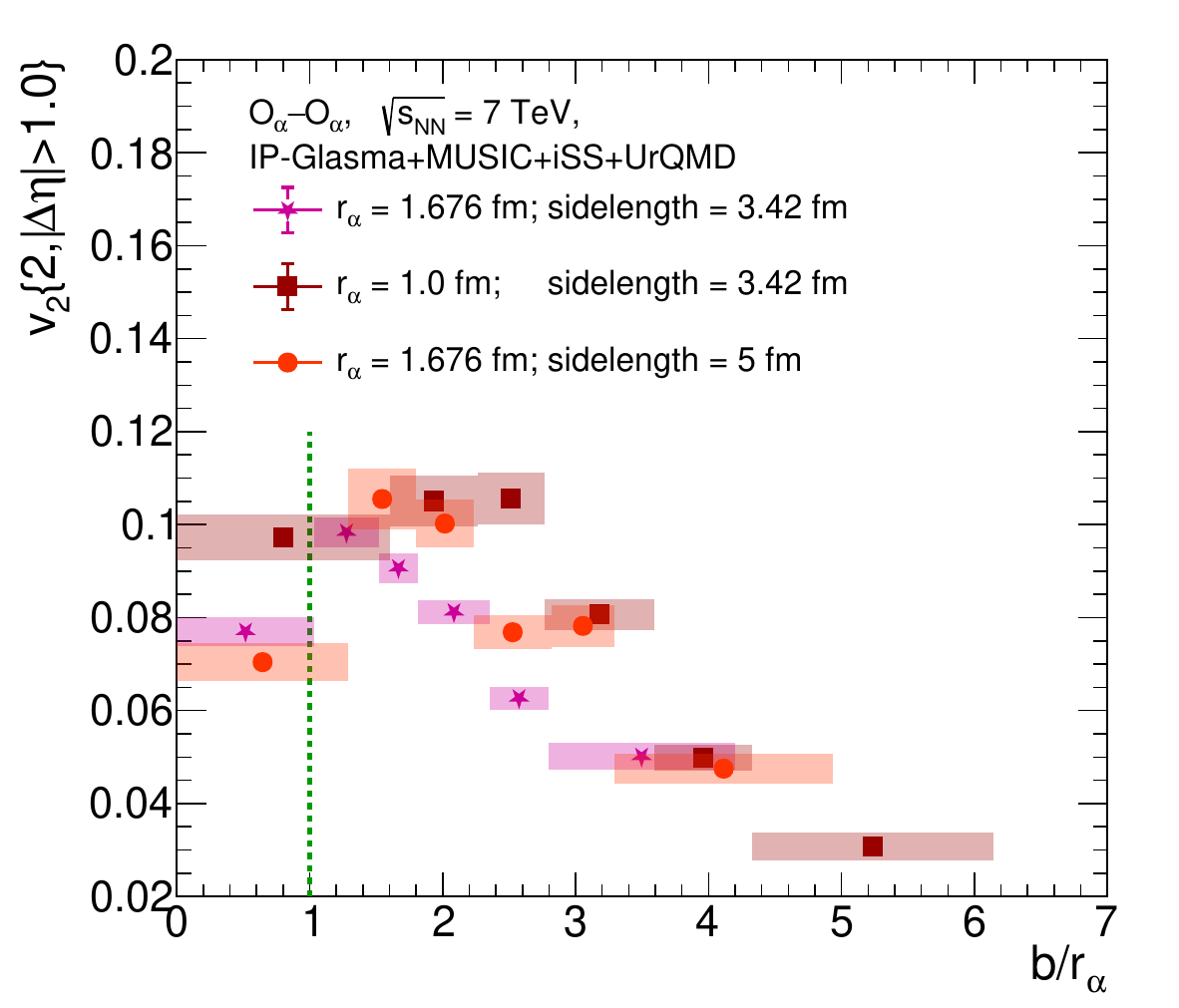}
\caption{$\langle\epsilon_{3}\rangle$ (left) and $v_{\rm 2}\{2, |\Delta\eta|>1.0\}$ (middle) as a function of centrality and $v_{\rm 2}\{2, |\Delta\eta|>1.0\}$ as a function of $\rm b/\rm r_{\alpha}$ (right) for different $\alpha$-cluster parametrizations of $^{16}\rm O$ nucleus for OO collisions at $\sqrt{s_{\rm NN}}=7$ TeV using IP-Glasma + MUSIC + iSS + UrQMD model. Estimated statistical uncertainties are well within the marker sizes or are incorporated in the error bands.}
\label{fig:Appndx1}
\end{figure*}

In Fig.~\ref{fig:ecc}, the enhanced value of $\langle\epsilon_{3}\rangle$ in the most central OO collisions with $\alpha$-clustered nuclear profile, stands out as a unique feature of $\alpha$-clustering in $^{16}\rm O$ nucleus. If one comes to the observations on $v_{2}$ in Figs.~\ref{fig:vn},~\ref{fig:vn_<Nch>} and~\ref{fig:vn_bralpha}, the important finding that distinguishes $\alpha$-cluster profile from that of Woods--Saxon is the peak-like structure present in the former for the centrality (or impact-parameter) dependent trend for $v_{2}$ in the 10-20\% class.

To test the consistency of these trends, in Fig.~\ref{fig:Appndx1}, $\langle\epsilon_{3}\rangle$ and $v_{2}$ as a function of centrality and $v_{2}$ as a function of $b/r_{\rm \alpha}$ are plotted for OO collisions, by varying the $\alpha$-cluster parametrizations inside the colliding $^{16}\rm O$ nuclei. The different trial cases attempted are explained below:
\begin{itemize}
    \item CASE 0: $r_{\rm \alpha} = 1.676$~fm and tetrahedral sidelength = 3.42~fm (default parametrization)
    \item CASE 1: $r_{\rm \alpha}$ is decreased to 1.0~fm. Tetrahedral sidelength remains unchanged.
    \item CASE 2: $r_{\rm \alpha}$ remains unchanged. Tetrahedral sidelength is increased to 5 fm.
\end{itemize}

In the left plot of Fig.~\ref{fig:Appndx1}, it can be seen that the rise in value of $\langle\epsilon_{3}\rangle$ in the most central OO collisions is present in all three cases of study. In fact, this rise is steeper in cases 1 and 2 than in case 0 (the default $\alpha$-cluster parametrization used in our work), which is due to the well-distinct non-overlapping clusters within the nucleus in cases 1 and 2\footnote{However, it is important to note that cases 1 and 2 are not realistic configurations as far as the nucleus under study is concerned and are useful only to validate the consistency of the trends in our results.}. This brings us to the conclusion that the increased value of $\langle\epsilon_{3}\rangle$ in most central OO collisions involving $\alpha$-clusters is indeed a green signal to the presence of grouping of nucleons into $\alpha$-clusters in the nuclei. The middle and right plots of Fig.~\ref{fig:Appndx1} are presented as a cross-check for the existence of the peak-like structure in the centrality or impact-parameter dependence of $v_{2}$. From these figures, it is to be understood that the peak-like feature in $v_{2}$ is a prominent one, in all cases considered. Once again, the magnitudes are slightly higher for cases 1 and 2 than for case 0. However, the peaks can be seen to get shifted towards increased impact-parameter classes for the unrealistic configurations (cases 1 and 2) in comparison to the realistic/default configuration. In short, the connection of cluster size to the elliptic flow still awaits validation with stable and natural clustered nuclear configurations, while the peak-like structure in the trend for $v_{2}$ in the slightly off-central OO collisions stands as a testable signature for $\alpha$-clustering in $^{16}\rm O$ nucleus.

\subsection{$\langle v_{2}\rangle$--$\epsilon_{2}$ correlation}
\label{AppendixB}

\begin{figure*}[ht!]
\centering
\includegraphics[scale=0.44]{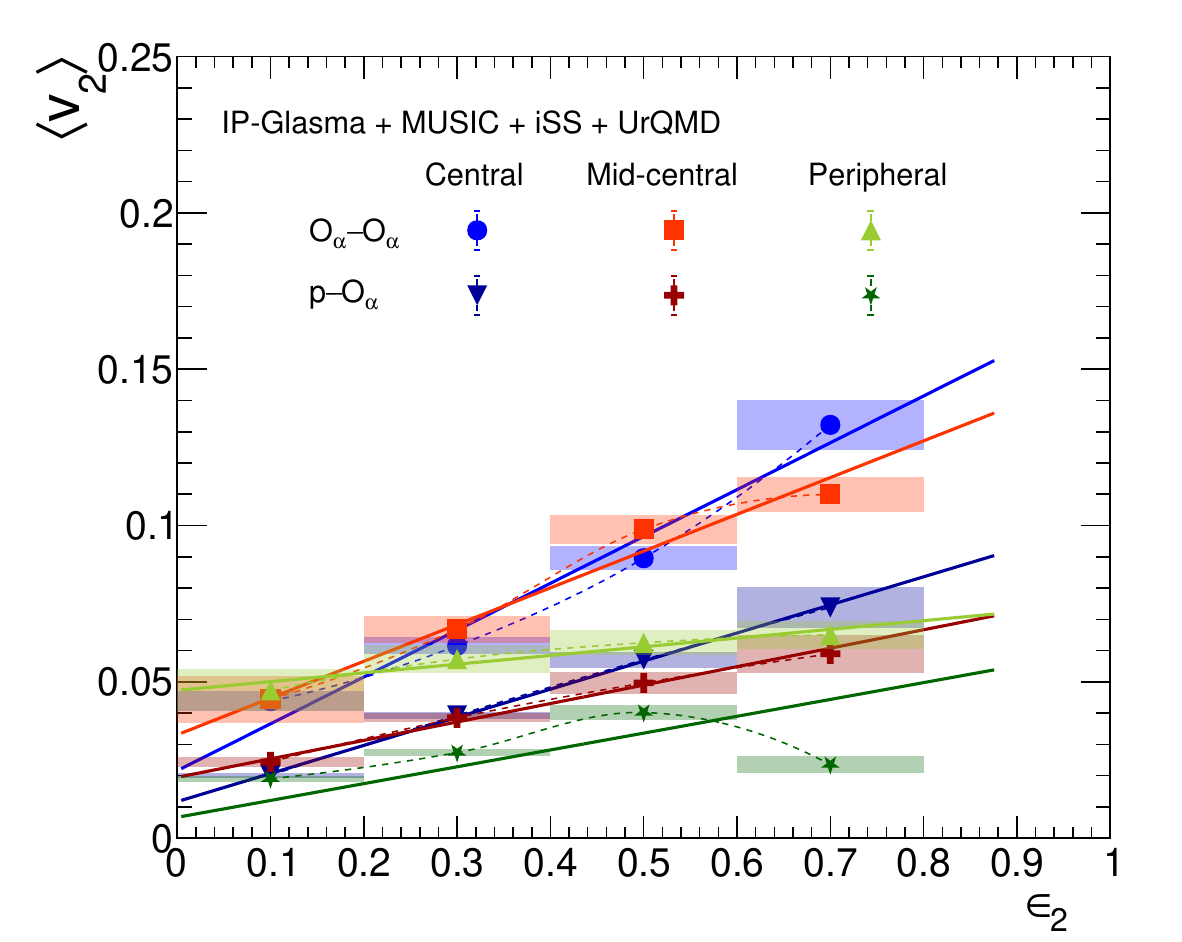}
\includegraphics[scale=0.44]{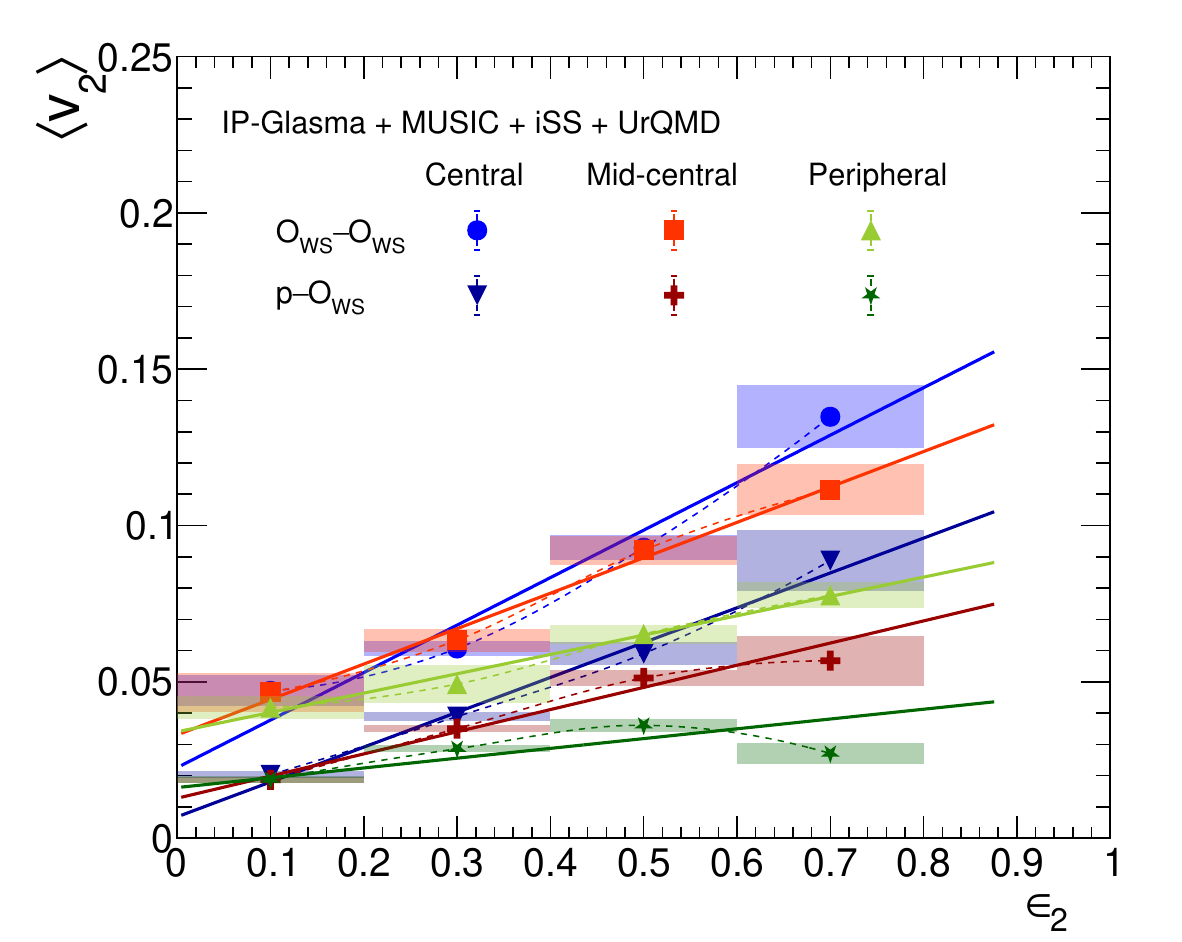}
\caption{$\langle v_{2}\rangle$ versus $\epsilon_{2}$ as a function of collision centrality classes for OO and pO collisions at $\sqrt{s_{\rm NN}}=$~7~TeV and 9.61~TeV respectively, for $\alpha$-cluster (left) and Woods--Saxon (right) nuclear density profiles of $^{16}\rm O$ nucleus, using IP-Glasma + MUSIC + iSS + UrQMD model. Error bands incorporate the estimated statistical uncertainties.}
\label{fig:e2v2LINEAR}
\end{figure*}

To a good approximation, elliptic flow is understood to be predominantly contributed by the linear response from the initial eccentricity, though deviations are found in peripheral collisions or towards higher values of $\epsilon_{2}$~\cite{Prasad:2025ezg,Alver:2010gr,Gardim:2011xv,Noronha-Hostler:2015dbi}. Hence, it is important to examine the overall validity of the linear scaling of eccentricity with elliptic flow \textit{i.e.} $v_{2}\approx k_{2}\epsilon_{2}$. With this intention, Fig.~\ref{fig:e2v2LINEAR} presents the correlation between $\langle v_{2} \rangle$ and $\epsilon_{2}$ (dotted lines) for OO and pO collisions (for both Woods--Saxon and $\alpha$-cluster nuclear density profiles of oxygen) for different classes of collision centrality. The solid lines are the corresponding straight-line fits. From the figures, it becomes evident that the hydrodynamic response (quantified by the parameter, $k_{2} = v_{2}/\epsilon_{2}$) is approximately linear in nature. However, the slope of the straight lines decreases as one moves from central to peripheral collisions, as well as when one moves from OO to pO collisions. This weaker response in low multiplicity environments is expected because of the lower lifetime of the systems formed in peripheral collisions and light-nuclei collisions. Additionally, a fluctuating system response is also hinted at, which becomes prominent towards the small collision systems and towards the peripheral collisions.


\end{document}